\documentclass[10pt,twocolumn,twoside] {IEEEtran}

\usepackage{amsbsy}			
\usepackage{amsmath}
\usepackage{amssymb}
\usepackage{amsfonts}
\usepackage{booktabs}
\usepackage{multirow}
\usepackage{latexsym}
\usepackage{amsthm}
\usepackage{ifthen}
\usepackage{tabularx}
\usepackage{csquotes}
\usepackage{lipsum}
\usepackage{bm}
\usepackage{algorithm}	
\usepackage{algorithmic}
\usepackage{graphicx,epstopdf}
\usepackage{caption}
\usepackage{subcaption}
\usepackage[normalem]{ulem}
\usepackage{xspace}
\usepackage{paralist}		
\usepackage{breqn}
\usepackage{cite}
\usepackage{color}
\usepackage[utf8]{inputenc}
\usepackage{textcomp}
\usepackage{arydshln}

\allowdisplaybreaks

\newtheorem{theorem}{Theorem}
\newtheorem{lemma}{Lemma}
\newtheorem{proposition}{Proposition}

\newtheorem{remark}{Remark}
\newtheorem{assumption}{Assumption}

\let\oldremark \remark
\renewcommand{\remark}{\oldremark\normalfont}

\newcommand{\bx}{{\bm x}}
\newcommand{\bv}{{\bm v}}
\newcommand{\bz}{{\bm z}}
\newcommand{\br}{{\bm r}}

\newcommand{\by}{{\bm y}}

\newcommand{\xf}{\widehat{\bx}_{i|i}}
\newcommand{\xp}{\widehat{\bx}_{i+1|i}}
\newcommand{\yf}{\widehat{\by}_{i|i}}
\newcommand{\yp}{\widehat{\by}_{i+1|i}}
\newcommand{\xpp}{\widehat{\bx}_{i|i-1}}
\newcommand{\ypp}{\widehat{\by}_{i|i-1}}

\newcommand{\ef}{{\bm{e}}_{i|i}}
\newcommand{\ep}{{\bm{e}}_{i+1|i}}
\newcommand{\epp}{{\bm{e}}_{i|i-1}}
\newcommand{\epsf}{{\bm{\epsilon}}_{i|i}}
\newcommand{\epsp}{{\bm{\epsilon}}_{i+1|i}}
\newcommand{\epspp}{{\bm{\epsilon}}_{i|i-1}}

\begin{document}

\bstctlcite{IEEEexample:BSTcontrol}

\title{\textit{Consensus+Innovations} Distributed Kalman Filter with Optimized Gains}

\author{Subhro~Das,~\IEEEmembership{Member,~IEEE,}
              and~Jos\'e~M. F.~Moura,~\IEEEmembership{Fellow,~IEEE}  
\thanks{Copyright (c) 2016 IEEE. Personal use of this material is permitted. However, permission to use this material for any other purposes must be obtained from the IEEE by sending a request to pubs-permissions@ieee.org. \bf{Published in IEEE Transactions on Signal Processing, October 12, 2016, DOI: 10.1109/TSP.2016.2617827}} 
\thanks{S. Das is at IBM T.J. Watson Research Center, Yorktown Heights, NY 10598, USA (ph: (914) 945-2089; e-mail: subhro.das@ibm.com). He performed the work when he was a PhD student at Department of Electrical and Computer Engineering, Carnegie Mellon University. } 
\thanks{J. M. F. Moura is with the Department of Electrical and Computer Engineering, Carnegie Mellon University, Pittsburgh, PA 15213 USA (ph: (412) 268-6341; e-mail: moura@ece.cmu.edu).} 
\thanks{This work was supported by NSF grant CCF1513936. The simulations were run in Microsoft Azure cloud supported by Microsoft Azure for Research Grant.}
}  

\maketitle

%
\begin{abstract}
In this paper, we address the distributed filtering and prediction of time-varying random fields represented by linear time-invariant (LTI) dynamical systems. The field is observed by a sparsely connected network of agents/sensors collaborating among themselves. We develop a Kalman filter type \textit{consensus$+$innovations} distributed linear estimator of the dynamic field termed as \textit{Consensus+Innovations} Kalman Filter. We analyze the convergence properties of this distributed estimator. We prove that the mean-squared error of the estimator asymptotically converges if the degree of instability of the field dynamics is within a pre-specified threshold defined as tracking capacity of the estimator. The tracking capacity is a function of the local observation models and the agent communication network. We design the optimal consensus and innovation gain matrices yielding distributed estimates with minimized mean-squared error. Through numerical evaluations, we show that, the distributed estimator with optimal gains converges faster and with approximately 3dB better mean-squared error performance than previous distributed estimators.
\end{abstract}

\begin{keywords}
Kalman filter, distributed estimation, multi-agent networks, distributed algorithms, consensus.
\end{keywords}


%
\section{Introduction}
\label{sec:intro}
For decades, the Kalman-Bucy filter \cite{kalman1961new, kalman1960new} has played a key role in estimation, detection, or prediction of time-varying noisy signals. The Kalman filter is found in a wide variety of applications ranging from problems in navigation to environmental studies, computer vision to bioengineering, signal processing to econometrics. More recently, algorithms inspired by the Kalman filter have been applied to estimate random fields monitored by a network of sensors. In these problems, we distinguish two distinct layers: (a) the physical layer of the time-varying random field; and (b) the cyber layer of sensors observing the field.

A centralized approach to field estimation poses several challenges. It requires that all sensors communicate their measurements to a centralized fusion center. This is fragile to central node failure and severely taxes computationally the fusion center. Moreover, it also requires excessive communication bandwidth to and from the fusion center. Hence, the centralized approach is inelastic to estimation of large-scale time-varying random fields, like, when estimating temperature, rainfall, or wind-speed over large geographical areas~\cite{de2011kalman},~\cite{battistelli2015distributed}.

In \cite{das2015TSP}, we proposed the {\it Distributed Information Kalman Filter} (DIKF) that is a distributed estimator of time-varying random fields consisting of two substructures. The first is the {\it Dynamic Consensus on Pseudo-Observations} (DCPO), a distributed estimator of the global average of the pseudo-observations (modified versions of the observations) of the agents. The second substructure uses these average estimates of pseudo-observations to estimate the time-varying random field. In this paper, we develop a distributed Kalman filter like estimator, the \textit{Consensus$+$Innovations} Kalman Filter (CIKF), that instead of using the pseudo-observations uses distributed estimates of the pseudo-state (modified version of a state) to estimate the field. We show how to design optimally the gain matrices of the CIKF. We prove that the CIKF converges in the mean-squared error (MSE) sense when the degree of instability of the dynamics of the random field is within the network tracking capacity \cite{khan2010connectivity}, a threshold determined by the cyber network connectivity and the local observation models.

\begin{figure*}
	\centering
	\begin{subfigure}[b]{0.45\textwidth}
		\includegraphics[width=\textwidth]{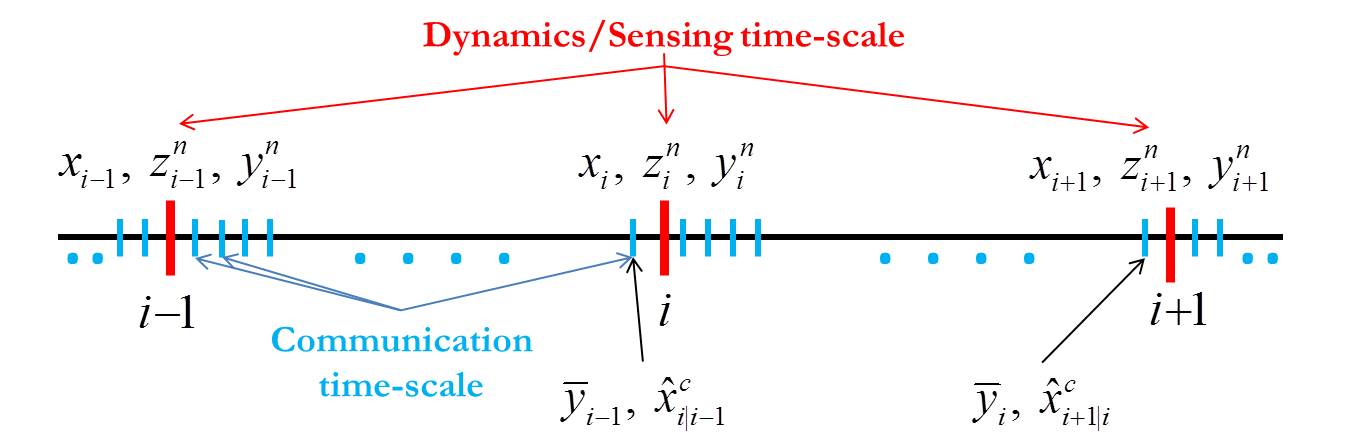}
		\caption{Two time-scale}
		\label{fig:twoTS}
	\end{subfigure}%
	~~~~
	\begin{subfigure}[b]{0.45\textwidth}
		\includegraphics[width=\textwidth]{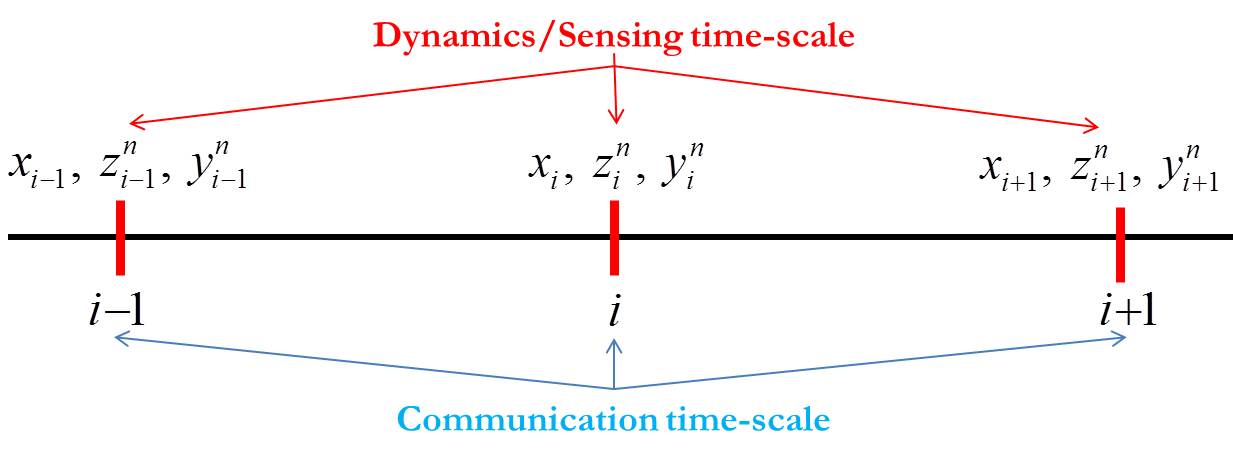}
		\caption{Single time-scale}
		\label{fig:singleTS}
	\end{subfigure}
	\caption{\small Time-scales of operation of dynamics, sensing and communications.}
	\vskip5pt
	\hrule
	\vskip-5pt
	\label{fig:references}
\end{figure*}

We review related prior research on distributed estimation of time-varying random fields. We classify prior work into two categories based on the time-scales of operation: (a) two time-scale and (b) single time-scale. In two-time scale distributed estimators, see Fig~\ref{fig:twoTS}, agents exchange their information multiple number of times between each dynamics/observations time-scale \cite{olfati2005distributed, olfati2007distributed, khan2008distributing, carli2008distributed, schizas2008consensus, ribeiro2010kalman, olfati2009kalman, cattivelli2010diffusion, casbeer2009distributed}, so that average consensus occurs between observations. In contrast, in single time-scale approaches \cite{khan2010connectivity, khan2011networked,khan2011coordinated,khan2013genericity, kar2011gossip, li2015distributed, kar2012distributed,kar2011convergence, martins2012augmented, karmoura-spm2013, das2015TSP}, the agents collaborate with their neighbors only once in between each dynamics/observation evolution. In other words, the dynamics, observation, and communication follow the same time scale as depicted in Fig~\ref{fig:singleTS}.

The two time-scale approach demands fast communication between agents. In most practical applications, this is not true. Further, references~~\cite{olfati2009kalman, cattivelli2010diffusion} developed distributed state estimators assuming local observability of the dynamic state in the physical layer. Such assumption is not feasible in large-scale systems. References~\cite{rao1991fully,ribeiro2006soi} assume a complete cyber network, which is not scalable.

We are interested in a single time-scale approach. We consider that the time-varying system is not locally observable at each agent, and we assume that the communication network in the cyber layer is sparsely connected. In the single time-scale category, references~\cite{kar2011gossip, li2015distributed} propose distributed Kalman filters where the agents communicate among themselves using the Gossip protocol \cite{dimakiskarmourarabbatscaglione-2010}. Although Gossip filters require very low communication bandwidth, their MSE is higher and their convergence rate is lower than the {\it consensus$+$innovations} based distributed approaches. References~\cite{kar2012distributed,kar2011convergence,karmoura-spm2013} introduced \textit{consensus$+$innovations} type distributed estimator for parameter (static state) estimation. This approach is extended to estimating time-varying random  states in \cite{khan2010connectivity, khan2011networked, khan2011coordinated, khan2013genericity}. Distributed {\it consensus$+$innovations} dynamic state estimators converge in MSE sense if their degree of instability of the state dynamics is below the network tracking capacity, see~\cite{khan2010connectivity}. In this paper, we derive the tracking capacity for the CIKF. Its tracking capacity is a function of the local observation models and of the agents communication connectivity.

The single time-scale \textit{consensus$+$innovations} distributed estimators that we introduced in~\cite{das2013ICASSP, das2013EUSIPCO, das2013Allerton, das2013Asilomar} run a companion filter to estimate the global average of the pseudo-innovations, a modified version of the innovations. The DIKF, we proposed in~\cite{das2015TSP}, uses averaged pseudo-observations (linearly transformed observations) rather than pseudo-innovations. In contrast, this paper uses estimates of the pseudo-state, a linear transformation of the dynamic state. In centralized information filter, the pseudo-state is directly available from the observations. In the distributed setting, since not all the observations are available to the local sensors, the CIKF has to distributedly estimate the pseudo-state through a consensus step. Using pseudo-state rather than pseudo-innovations~\cite{das2013ICASSP, das2013EUSIPCO, das2013Allerton, das2013Asilomar} or pseudo-observations~\cite{das2015TSP} leads to significant better performance as we show here. The MSE of the CIKF is lower than that of the distributed estimators in~\cite{kar2011gossip,li2015distributed,das2013ICASSP, das2013EUSIPCO, das2013Allerton, das2013Asilomar, das2015TSP}.

Developing distributed estimators for time-varying random fields (\enquote{distributed Kalman filter}) has gained considerable attention over the last few years. The goal has been to achieve MSE performance as close as possible to the optimal centralized Kalman filter. We show in this paper that the CIKF converges to a bounded MSE solution requiring minimal assumptions, namely global detectability and connected network, and not requiring the additional distributed observability assumption, as needed by the DIKF. The distributed optimal time-varying field estimation is, in general, NP-hard. The distributed parameter estimation~\cite{kar2012distributed, kar2014asymptotically} have shown asymptotic optimality, in the sense that the distributed parameter estimator is asymptotically unbiased, consistent, and efficient converging at the same rate as the centralized optimal estimator. However, estimation of time-varying random fields adds another degree of complication since, while information diffuses through the network, the field itself evolves. So, this lag causes a gap in performance between distributed and centralized field estimators. Numerical simulations show that the proposed CIKF improves the performance by 3dB over the DIKF, reducing by half the gap to the centralized (optimal) Kalman filter, while showing a faster convergence rate than the DIKF. These improvements significantly distinguish the CIKF from the DIKF.

The rest of the paper is organized as follows. We describe the three aspects of the problem setup in Section~\ref{sec:problem}. Section~\ref{sec:solution} introduces the pseudo-state and presents the proposed optimal gain distributed Kalman filter (CIKF). In Section~\ref{sec:error}, we analyze the dynamics of the error processes and their covariances. Section~\ref{sec:capacity} includes the analysis of the tracking capacity of the proposed CIKF. We design the optimal gain matrices to obtain distributed estimates in Section~\ref{sec:gain}. Numerical simulations are in Section~\ref{sec:simulation}. We present the concluding remarks in Section~\ref{sec:conclusion}. Proofs of the proposition, lemmas and theorems are in the Appendix~\ref{app:proof_pstate}, Appendix~\ref{app:proof_error}-\ref{app:proof_uncorrelation}, and Appendix~\ref{app:proof_OKDF}-\ref{app:proof_gain}, respectively.

%
\section{System, Observation, and Communication}
\label{sec:problem}
\begin{figure*}
	\centering
	\begin{subfigure}[b]{0.32\textwidth}
		\includegraphics[width=\textwidth]{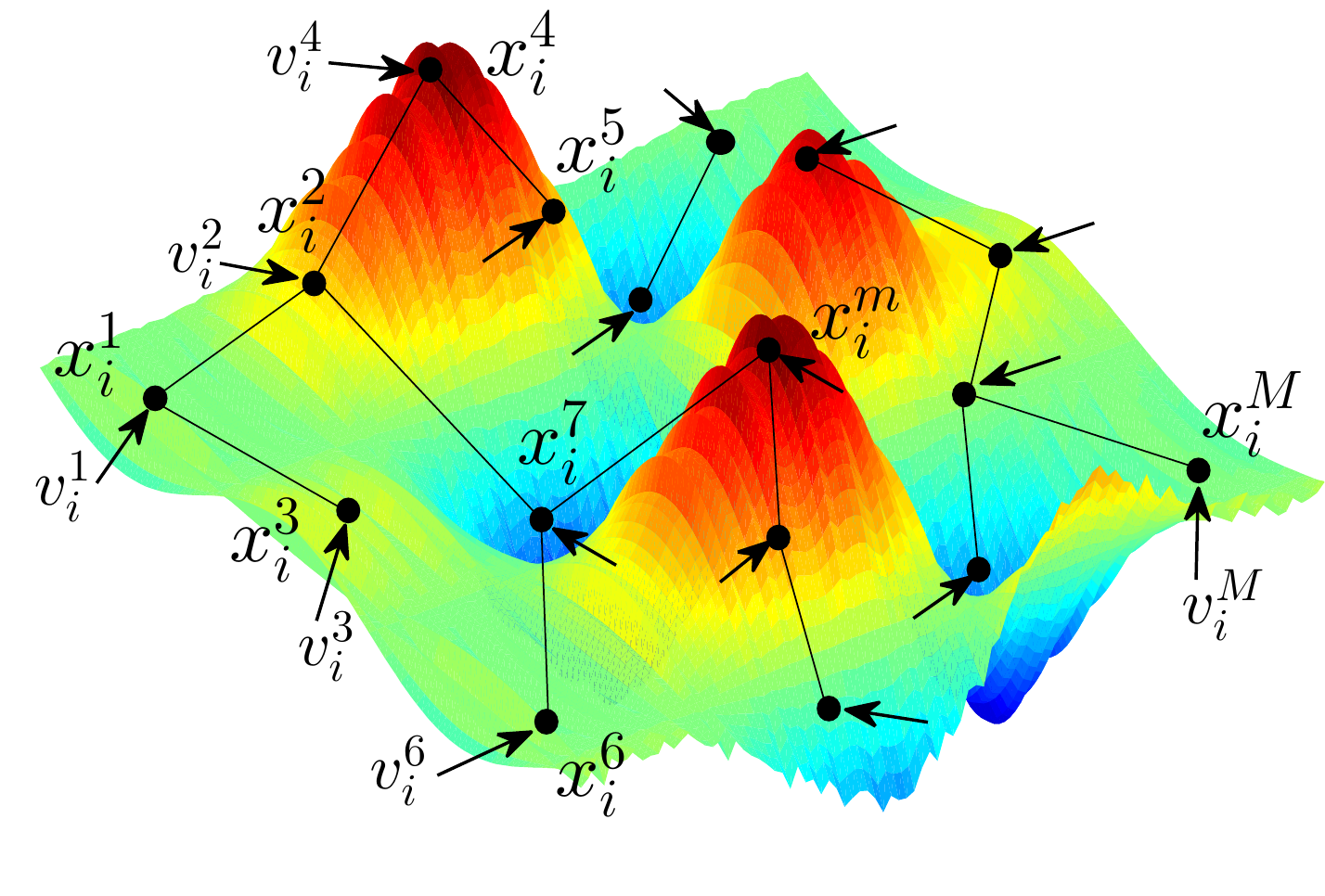}
		\caption{Dynamical system: random field}
		\label{fig:dynamics}
	\end{subfigure}%
	~
	\begin{subfigure}[b]{0.32\textwidth}
		\includegraphics[width=\textwidth]{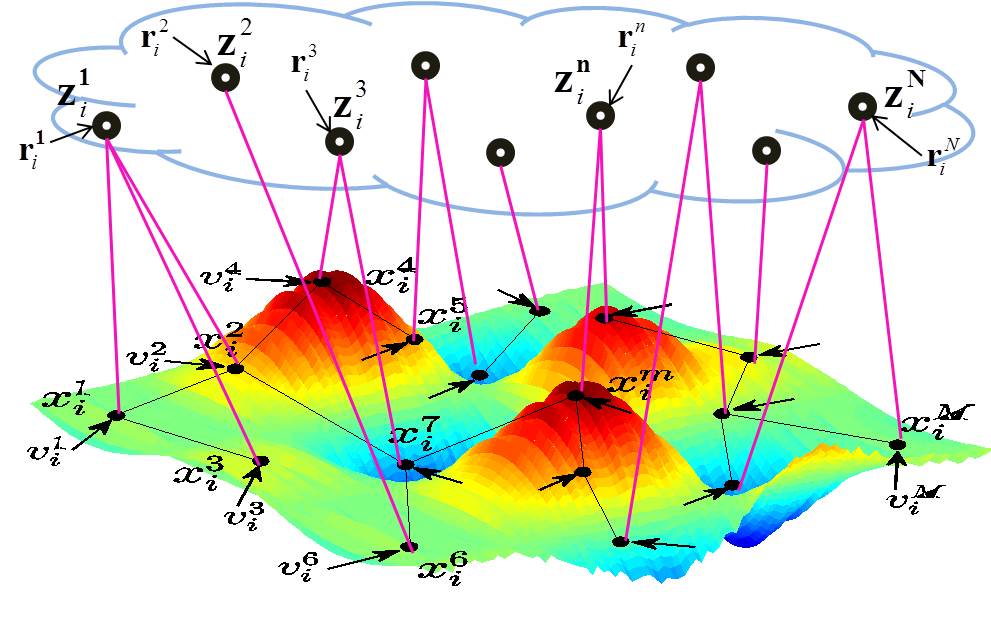}
		\caption{Agent network: local measurements}
		\label{fig:observations}
	\end{subfigure}%
	~
	\begin{subfigure}[b]{0.32\textwidth}
		\includegraphics[width=\textwidth]{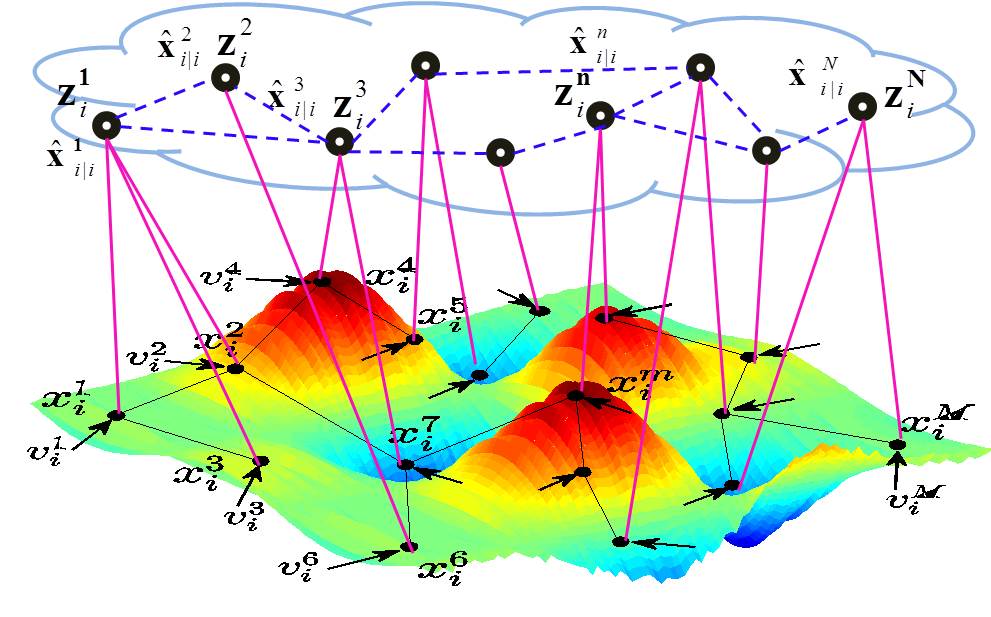}
		\caption{Agent network: local communications}
		\label{fig:communication}
	\end{subfigure}
	\caption{\small The evolution of a random field as a discrete-time linear dynamical system. The field is observed by a network of~$N$ sensors (agents). The agents collaborate locally via agent communication network.}
	\vskip5pt
	\hrule
	\vskip-5pt
	\label{fig:problem1}
\end{figure*}
The distributed estimation framework consists of three components: dynamical system, local observation, and neighborhood communications. These three parts include two layers: the physical layer and the cyber layer. For the sake of simplicity, we motivate the model with the example of a time-varying temperature field over a large geographical area monitored by a sensor network.

\subsection{Physical layer: dynamical system}
\label{subsec:system}
Consider a time-varying temperature field distributed over a large geographical area, as shown in Fig~\ref{fig:dynamics}. A first-order approximation and discretization of the temperature field provide spatio-temporal discretized temperature variables $x^j_i, j = 1, \cdots, M$, of $M$~sites at discrete time indexes $i = 1, \cdots, \;$. We stack the $M$ field variables in a temperature state vector $\bx_i = \begin{bmatrix} x^1_i, \cdots, x^M_i \end{bmatrix}^T \in \mathbb{R}^{M}$. The evolution of the time-varying temperature field, $\bx_i$, can be represented by a linear time-invariant (LTI) dynamical system\footnote{Even though the random field~$\bx_i$ is time-varying, the system is time-invariant because the field dynamics matrix~$A$ is time-invariant. For a time-varying dynamical system, the field dynamics matrix~$A_i$ would be a function of time.}
\begin{align}
\label{eqn:sys}
\bx_{i+1} = A \bx_i + \bv_i,
\end{align}
where the first-order dynamics matrix $A \in \mathbb{R}^{M \times M}$ contains the coupling effects between the $M$ temperature variables, and the residual $\bv_i = \begin{bmatrix} v^1_i, \cdots, v^M_i \end{bmatrix}^T \in \mathbb{R}^{M}$ is the system noise driving the dynamical temperature field\footnote{The temperature field is a time-varying random field. Although the field is time-varying, it is generated by a linear time-invariant system with perturbations.}. At each of the $M$ sites, the input noise $v^j_i, j = 1, \cdots, M$ accounts for the deviations in the temperature after the overall field dynamics. The field dynamics $A$ incorporates the sparsity pattern and connectivity of the physical layer consisting of the dynamical system~\eqref{eqn:sys}.
%
\subsection{Cyber layer: local observations}
\label{subsec:obs}
The physical layer consisting of the field dynamics~\eqref{eqn:sys} is observed by a cyber layer consisting of a network of~$N$ agents (sensors). In Fig~\ref{fig:observations}, we see that there are~$N$ sensors, where each sensor observes the temperatures of only a few sites. Denote the number of sites observed by sensor~$n$ by~$M_n, M_n \ll M$, and its measurements at time~$i$ by~$z_i^n \in \mathbb{R}^{M_n}$. The observations of the agents in the cyber layer can be represented by a linear model:
\begin{align}
\label{eqn:obs}
\bz_i^n = H_n \bx_i + \br_i^n,  \qquad \qquad n =1, \hdots, N,
\end{align}
where, the observation matrix~$H_n \in \mathbb{R}^{M_n \times M}$ contains the observation pattern and strength information, and the observation noise $r_i^n \in \mathbb{R}^{M_n}$ reflects the inaccuracy in measurements due to sensor precision, high frequency fluctuations in temperature, and other unavoidable constraints.

To illustrate, consider that sensor~1 is observing the temperature of 3~sites $\{ x^1_i, x^2_i, x^3_i \}$, i.e., $M_1 = 3$. An example snapshot of the observation model of agent~$n$ is
\begin{small}
\begin{align*}
\underbrace{\begin{bmatrix} z^{11}_i \\ z^{12}_i \\ z^{13}_i \end{bmatrix}}_{\bz^1_i} &=
\!\! \underbrace{\begin{bmatrix} 1 & 0 & 0 & \hdots & 0 \\ 0 & 5 & 0 & \hdots & 0 \\ 0 & 0 & 6 & \hdots & 0 \end{bmatrix}}_{H_1}
\!\! \underbrace{\begin{bmatrix} x^1_i \\ \vdots \\ x^{\scriptsize{M}}_i \end{bmatrix}}_{\bx_i}
\!+\! \underbrace{\begin{bmatrix} r^{11}_i \\ r^{12}_i \\ r^{13}_i \end{bmatrix}}_{\br^1_i}
= \begin{bmatrix} x^1_i \\ 5x^2_i \\ 6x^3_i \end{bmatrix} \!+\! \begin{bmatrix} r^{11}_i \\ r^{12}_i \\ r^{13}_i \end{bmatrix}.
\end{align*}
\end{small}
Now, for the ease of analysis, we aggregate the noisy local temperature measurements, $\bz^1_i, \cdots \bz^N_i$, of all the~$N$ agents in a global observation vector,~$\bz_i \in\mathbb{R}^{{\sum_{n=1}^{N} M_n}}$:
\begin{eqnarray}
\label{eqn:obs_global}
\underbrace{\begin{bmatrix} \bz_i^1\\  \vdots \\ \bz_i^N\end{bmatrix}}_{\bz_i} &=&
\underbrace{\begin{bmatrix}H_1 \\ \vdots \\ H_N \end{bmatrix}}_{H} \bx_i +
\underbrace{\begin{bmatrix} \br_i^1 \\ \vdots \\ \br_i^N\end{bmatrix}}_{\br_i},
\end{eqnarray}
where, the global observation matrix is $H\in \mathbb{R}^{\sum_{n=1}^N M_n \times M}$ and the stacked measurement noise is $r_i\in\mathbb{R}^{{\sum_{n=1}^{N} M_n}}$. Note that, in general, the temperature measurement model is non-linear. For non-linear cases, refer to distributed particle filter in~\cite{mohammadi2015distributed} and the references cited therein. Here we perform a first-order approximation to obtain a linear observation sequence.

%
\subsection{Cyber layer: neighborhood communication}
\label{subsec:comm}
In the cyber layer, the agents exchange their temperature readings or current estimates with their neighbors. In many applications, to reduce communications costs, neighbors communicate only with their geographically nearest agents as shown in Fig~\ref{fig:communication}.

Formally, the agent communication network is defined by a simple (no self-loops nor multiple edges), undirected, connected graph $\mathcal{G = (V,E)}$, where $\mathcal{V} $ is the set of sensors (nodes or agents) and  $\mathcal{E}$ is the set of local communication channels (edges or links) among the agents. The open~$\Omega_n$ and closed~$\overline{\Omega}_n$ neighborhoods of agent~$n$ are:
\begin{eqnarray}
\label{eqn:comm}
\Omega_n &= \{l|(n,l) \in  \mathcal{E}\}. \nonumber \\
\overline{\Omega}_n &=  \{n\} \cup \{l|(n,l) \in  \mathcal{E}\}. \nonumber
\end{eqnarray}
In Fig~\ref{fig:communication} the open and closed neighborhoods of agent~1 are $\Omega_1 = \{2,3\}$ and $\overline{\Omega}_1 = \{1, 2,3\}$, respectively. The Laplacian matrix of $\mathcal{G}$ is denoted by $L$. The eigenvalues of the positive semi-definite matrix $L$ are ordered as $ 0 = \lambda_1(L) \leq \lambda_2(L) \leq . . . \leq \lambda_N(L)$. For details on graphs refer to \cite{chung1997spectral}. The communication network is sparse and time-invariant.
%

\subsection{Modeling assumptions}
\label{subsec:assm}
We make the following assumptions.
\begin{assumption}[Gaussian processes]
	\label{assm:gauss}
 	The system noise,~$\bv_i$, the observation noise,~$\br_i$, and the initial condition of the system,~$\bx_0$, are Gaussian sequences, with
 	\begin{align*}
 	{\bv_i \sim \mathcal{N}(0,V)}, \;\;
 	{\br^n_i \sim \mathcal{N}(0,R_n)}, \;\;
 	\bx_0 \sim \mathcal{N}(\bar{\bx}_0,\Sigma_0),
 	\end{align*}
 	where, $V \in \mathbb{R}^{{M} \times {M}} $, $R_n \in \mathbb{R}^{{M_n} \times {M_n}}$ and $\Sigma_0 \in \mathbb{R}^{{M} \times {M}}$ are the corresponding covariance matrices. The noise covariance matrix~$R$ of the global noise vector~$\bv_i$ in~\eqref{eqn:obs_global} is block-diagonal, i.e., $R=\textrm{blockdiag} \{R_1, \ldots , R_N \}$,  and positive-definite, i.e., $R > 0$.
\end{assumption}
\begin{assumption}[Uncorrelated sequences]
	\label{assm:uncorrelated}
	The system noise, the observation noise, and the initial condition: $\{ \{\bv_i\}_i, \{\br_i\}_i, \bx_0\}_{i \geq 0}$ are uncorrelated random vector sequences.
\end{assumption}
\begin{assumption}[Prior information]
	\label{assm:info}
	Each agent in the cyber layer knows the system dynamics model, $A$ and $V$, the initial condition statistics, $\bar x_0$ and $\Sigma_0$, the parameters of the observation model, $H$ and $R$, and the communication network model, $\mathcal{G}$.
\end{assumption}

In large-scale system applications, the dynamics, observation, and network Laplacian matrices, $A$, $H_n$, and $L$, are sparse, with $M_n \ll M$, and the agents communicate with only a few of their neighbors, $|\Omega_n| \ll N, \forall n$. In the dynamics~\eqref{eqn:sys} and observations~\eqref{eqn:obs}, we assume that there is no deterministic input. The results are readily extended if there is a known deterministic input.

%
\subsection{Centralized Information filter}
\label{subsec:CvsD}
\begin{figure}[!t]
	\centering
	\includegraphics[width=0.4\textwidth]{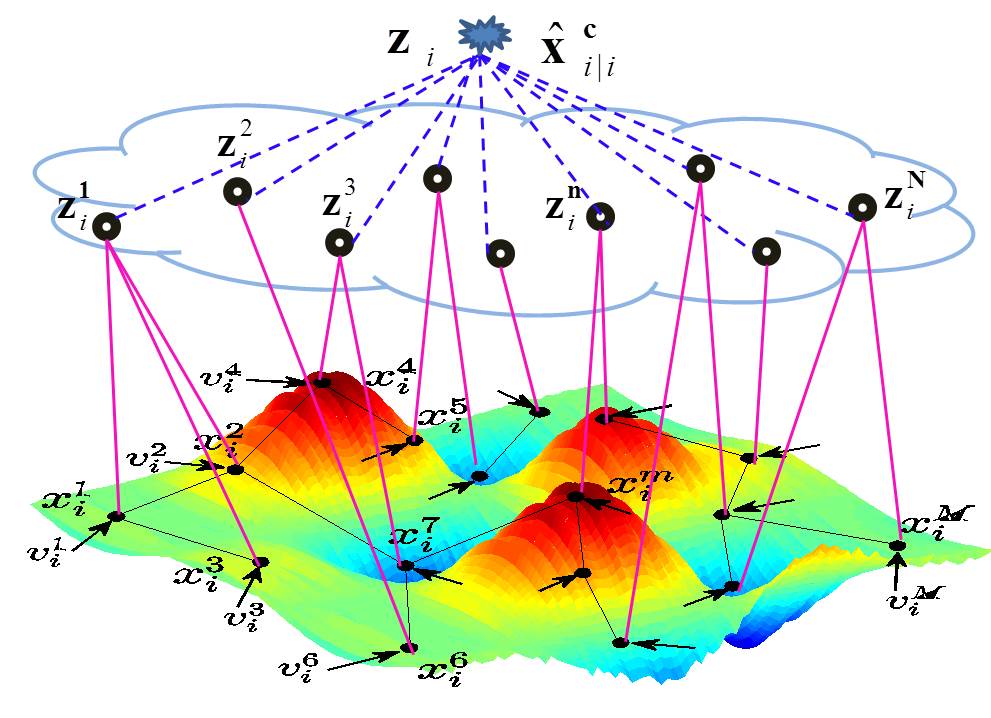}
	\caption{Centralized estimator}
	\label{fig:centralized}
\end{figure}

Although not practical in the context of the problem we study, we use the centralized information filter to benchmark our results on distributed estimator. In a centralized scheme, all the agents in the cyber layer communicate their measurements to a central fusion center, as depicted in Fig~\ref{fig:centralized}. The fusion center performs all needed computation tasks. Refer to \cite{das2015TSP, anderson2012optimal} for the filter, gain, and error equations of the centralized information filter.

%
\section{Distributed filtering and prediction}
\label{sec:solution}
This section considers our single time-scale distributed solution. We start with the introduction and derivation of the key components of our distributed estimator and then present our distributed field estimator.
%
\subsection{Pseudo-state model}
\label{subsec:pstate}
In a centralized information filter~\cite{anderson2012optimal}, all the observations are converted into pseudo-observations~\cite{das2015TSP} to obtain the optimal estimates. Following~\eqref{eqn:obs}, the pseudo-observation~$\widetilde{\bz}^n_i$ of agent~$n$ is
\begin{align}
\label{eqn:pseudo_obs}
\widetilde{\bz}^n_i &= H^T_n R_n^{-1} \bz^n_i =  \overline{H}_n \bx_i + H^T_n R_n^{-1} \br^n_i, \quad \\
\label{eqn:p_obs_H}
\text{where,} \qquad \overline{H}_n &= H^T_n R_n^{-1} H_n.
\end{align}
The centralized information filter computes the sum, $\overline{\bz}_i$ of all the pseudo-observations
\begin{align}
\label{eqn:avg_pseudo_obs}
\overline{\bz}_i &= \sum_{n=1}^{N}\widetilde{\bz}^n_i = G\bx_i + H^T R^{-1} \br_i \qquad   \\
\label{eqn:G}
\text{where,} \qquad \qquad G &= \sum_{n=1}^{N} H^T_n R_n^{-1} H_n = \sum_{n=1}^{N} \overline{H}_n .
\end{align}
The aggregated pseudo-observation, $\overline{\bz}_i$, is the key term in the centralized filter. It provides the innovations term in the filter updates enabling the filter to converge with minimum MSE estimates. However, in the distributed solution, each agent~$n$ does not have access to all the pseudo-observations; instead it can only communicate with its neighbors. To address this issue, in~\cite{das2015TSP} we introduced a dynamic consensus algorithm to compute the distributed estimates of the averaged pseudo-observations, $\overline{\bz}_i$, at each agent. In~\eqref{eqn:avg_pseudo_obs}, we note that the crucial term is $G\bx_i$ which carries the information of the dynamic state,~$\bx_i$; and the second term in~$\overline{\bz}_i$ in~\eqref{eqn:avg_pseudo_obs} is noise. We refer to it as the pseudo-state, $\by_i$,
\begin{align}
\label{eqn:pstate}
\by_i = G\bx_i.
\end{align}
The pseudo-state, $\by_i$, is also a random field whose time dynamics can be represented by a discrete-time linear dynamical system. The pseudo-observations~$\widetilde{\bz}^n_i$ are its linear measurements. We summarize the state-space model for the pseudo-state in the following proposition. See Appendix~\ref{app:proof_pstate} for the details of the proof.
\begin{proposition}
\label{prop:state_space}
The dynamics and observations of the pseudo-state $\by_i$ are:
\begin{align}
\label{eqn:pstate_dynamics}
\by_{i+1} &= \widetilde{A}\by_i + G\bv_i + \check{A}\bx_i \\
\label{eqn:pstate_obs}
\widetilde{\bz}^n_i &= \widetilde{H}_n\by_i + H^T_n R_n^{-1} \br^n_i + \check{H}_n \bx_i.
\end{align}
The pseudo-dynamics matrix $\widetilde{A}$, pseudo-observations matrix $\widetilde{H}_n$, and the matrices $\check{A}, \check{H}_n$, and $\widetilde{I}$ at agent~$n$ are:
\begin{align}
\label{eqn:p_dynamics_mat}
\widetilde{A} &= G A G^{\dagger} \\
\label{eqn:p_obs_mat}
\widetilde{H}_n &= H^T_n R_n^{-1} H_n G^{\dagger} = \overline{H}_n G^{\dagger}\\
\label{eqn:A_ch}
\check{A} &= GA\widetilde{I} \\
\label{eqn:Hn_ch}
\check{H}_n &= H^T_n R_n^{-1} H_n \widetilde{I} \\
\label{eqn:p_identity_mat}
\widetilde{I} &= I - G^{\dagger}G,
\end{align}
where, $G^{\dagger}$ denotes the Moore-Penrose pseudo-inverse of G.
\end{proposition}
In~\cite{das2015TSP}, the distributed information Kalman filter (DIKF) assumes distributed observability, i.e., it considers the case where $G$ is invertible. Under this assumption, $G^{\dagger} = G^{-1}$ and $\widetilde{I} = 0$. In this paper we relax the requirement of invertibility of $G$, proposing a distributed estimator for general dynamics-observation models under the assumption of global detectability. In most cases~$\widetilde{I}$ is low-rank.
In~\eqref{eqn:pstate_dynamics}, the term~$\left( G\bv_i~+~\check{A}\bx_i = \xi_i, \text{say} \right)$ can be interpreted as the pseudo-state input noise, which follows Gauss dynamics
\begin{align*}
\xi_i &\sim \mathcal{N} \left( \check{A} \overline{\bx}_i \; , \; GVG + \check{A} \Sigma_i \check{A}^T\right) \qquad\qquad \\
\text{where,} \qquad \Sigma_{i} &= \mathbb{E}\left[ (\bx_i - \overline{\bx}_i)(\bx_i - \overline{\bx}_i)^T \right].
\end{align*}
Similarly, in~\eqref{eqn:pstate_obs}, the term~$\big( \delta^n_i = H^T_nR_n^{-1}\br^n_i + \check{H}_n \bx_i \big)$ is the pseudo-state observation noise at agent~$n$, which is Gaussian
\begin{align*}
\delta^n_i \sim \mathcal{N} \Big( \check{H}_n \overline{\bx}_i , \overline{H}_n + \check{H}_n \Sigma_{i} \check{H}_n \Big).
\end{align*}
For the ease of analysis, we express the pseudo-state observation model in vector form by~$\widetilde{\bz}_i \in\mathbb{R}^{{\sum_{n=1}^{N} M_n}}$, the aggregate of the noisy local temperature measurements, $\widetilde{\bz}^1_i, \cdots \widetilde{\bz}^N_i$, of all the agents,
\begin{eqnarray}
\label{eqn:pobs_global}
\underbrace{\begin{bmatrix} \widetilde{\bz}_i^1\\  \vdots \\ \widetilde{\bz}_i^N\end{bmatrix}}_{\widetilde{\bz}_i} &=&
\underbrace{\begin{bmatrix} \widetilde{H}_1 \\ \vdots \\ \widetilde{H}_N \end{bmatrix}}_{\widetilde{H}} \by_i +  D_{H}^T R^{-1} \br_i +  \underbrace{\begin{bmatrix} \check{H}_1 \\ \vdots \\ \check{H}_N \end{bmatrix}}_{\check{H}} \bx_i,
\end{eqnarray}
where, the matrices $\widetilde{H}, \check{H} \in \mathbb{R}^{NM \times M}$, and, $D_{H} = \text{blockdiag} \{H_1, \cdots, H_N \}$.

We have established in~\eqref{eqn:pstate_dynamics}-\eqref{eqn:pstate_obs} the state-space dynamics and observations model of the pseudo-state,~$\by_i$. The structure of~\eqref{eqn:pstate_dynamics}-\eqref{eqn:pstate_obs} is similar to the dynamics and observations model \eqref{eqn:sys}-\eqref{eqn:obs} of the random field,~$\bx_i$. In the following subsection, we develop a distributed estimator with optimized gains to obtain unbiased estimates of the pseudo-state,~$\by_i$, and of the state,~$\bx_i$, at each agent with minimized mean-squared error. The pseudo-state $\by_i$ is a noise-reduced version of the sum $\overline{\bz_i}$ of the local pseudo-observations. We demonstrate in Section~\ref{sec:simulation} that, by sharing distributed estimates of pseudo-states, the CIKF achieves 3dB better MSE error performance than the DIKF~\cite{das2015TSP}, which shares distributed estimates of the summed pseudo-observations $\overline{\bz}_i$.

\subsection{\textit{Consensus$+$Innovations} Kalman Filter (CIKF)}
\label{subsec:CIKF}
At time~$i$, denote the $n^{\text{th}}$ agent's distributed filter and prediction estimates of the state~$\bx_i$ by $\xf^n$ and $\xp^n$ respectively. Similarly, its distributed filter and prediction estimates of the pseudo-state~$\by_i$ are denoted by $\yf^n$ and $\yp^n$. At any time~$i$, each agent~$n$ has access to its own pseudo-observation~$\widetilde{z}^n_i$ and receives the prediction state pseudo-state estimates, $\ypp^l$, $l \in \Omega_n$, of its neighbors at the previous time~$i-1$. Under this setup, the minimized MSE filter and prediction estimates are the conditional means,
\begin{align}
\label{eqn:MMSE_yf}
&\yf^n \!=\! \mathbb{E} \!\left[ \by_i \;|\; \widetilde{\bz}^n_i, \{ \ypp^l \}_{l \in \overline{\Omega}_n} \right] \\
\label{eqn:MMSE_xf}
&\xf^n \!=\! \mathbb{E} \left[ \bx_i \;|\; \yf^n \right] \\
\label{eqn:MMSE_yp}
&\!\!\!\!\! \yp^n \!=\! \mathbb{E} \left[ \by_{i+1} \;|\; \widetilde{\bz}^n_i, \{ \ypp^l \}_{l \in \overline{\Omega}_n} \right] \\
\label{eqn:MMSE_xp}
&\!\!\!\!\! \xp^n \!=\! \mathbb{E} \left[ \bx_{i\!+\!1} \;|\; \yf^n \right].
\end{align}
In~\eqref{eqn:MMSE_yf}, $\yf^n$ is the filtered estimate of the pseudo-state~$\by_i$ given all the pseudo-observations available at agent~$n$ up to time~$i$ including those of its neighbors. By the principle of recursive linear estimation, instead of storing all the pseudo-observations $ \{ \{ \widetilde{\bz}^n_t \}_{t = 0, \cdots, i}, \{ \widetilde{\bz}^{n_1}_t \}^{n_1 \in \Omega_n}_{t = 0, \cdots, i-1}, \{ \widetilde{\bz}^{n_2}_t \}^{n_2 \in \Omega_{n1}, \forall n_1}_{t = 0, \cdots, i-2}, \cdots \}$, we need only the current pseudo-observation~$\widetilde{\bz}^n_i$ and the pseudo-state estimates (including those of its neighbors) from the previous time instant $\{ \ypp^l \}_{l \in \Omega_n}$. The distributed pseudo-state estimate, $\yf^n$, is the expectation of the pseudo-state, $\by_i$, conditioned on its own pseudo-observation~$\widetilde{\bz}^n_i$ and the prior pseudo-state estimates, $\ypp^l, l \in \overline{\Omega}_n,$ received from its neighbors. In this paper, we formulate the distributed field estimate, $\xf^n$, as the expectation of the field, $\bx_i$, conditioned on the distributed pseudo-state estimate $\yf^n$. The distributed field estimates, $\xf^n$, are not optimal given all observations, but they are still optimal given only the pseudo-state estimates $\yf^n$. The filtered estimate~$\xf^n$ of~$\bx_i$ in~\eqref{eqn:MMSE_xf} depends on the current pseudo-state filtered estimate~$\yf^n$. Similarly, the prediction estimates~$\yp^n$ and~$\xp^n$ of~$\by_{i+1}$ and~$\bx_{i+1}$ in~\eqref{eqn:MMSE_yp}-\eqref{eqn:MMSE_xp}, respectively, are conditioned on the corresponding available quantities up to time~$i$.
\begin{theorem}
	\label{thm:OKDF}
	The iterative updates to compute the distributed filtered estimates with optimized gains in~\eqref{eqn:MMSE_yf}-\eqref{eqn:MMSE_xf} are:
	\begin{align}		
		\!\!\! \yf^n \!=\! \ypp^n &\!+\! \sum_{l \in \Omega_n} \underbrace{ \! B^{nl}_i \! \left( \ypp^l \!-\! \ypp^n \right) }_{\text{Consensus}} \nonumber \\
		\label{eqn:CIKF_yf}
			& \!+\! B^{nn}_i  \underbrace{ \left( \widetilde{\bz}^n_i \!-\! \left(\widetilde{H}_n \ypp^n \!+\! \check{H}_n \xpp^n \right) \right) }_{\text{Innovations}}, \\
		\label{eqn:CIKF_xf}	
		\!\!\! \xf^n \!=\! \xpp^n & \!+\! K^{n}_i \underbrace{ \left( \yf^n \!-\! G \xpp^n \right) }_{\text{Innovations}}
	\end{align}
	where, pseudo-state gain block matrix, $B_i \in \mathbb{R}^{MN \times MN}$, is $ B_i = \begin{bmatrix} B^{nl}_i	\end{bmatrix}_{n = 1, \cdots, N}^{l = 1, \cdots, N}, B^{nl}_i=0 \; \text{if} \; l \notin \Omega_n$. The state gain block-diagonal matrix is $K_i = \text{blockdiag}\{K^{1}_i, \cdots, K^{N}_i\}, K^{n}_i \in~\mathbb{R}^{M \times M}$. The optimized MSE prediction estimates in~\eqref{eqn:MMSE_yp}-\eqref{eqn:MMSE_xp} are:
	\begin{align}
		\label{eqn:CIKF_yp}
		\yp^n &= \widetilde{A}\yf^n + \check{A}\xf^n, \\
		\label{eqn:CIKF_xp}
		\xp^n &= A \xf^n.
	\end{align}
\end{theorem}
The proof of Theorem~\ref{thm:OKDF} is presented in Appendix~\ref{app:proof_OKDF}. The update equations reflect the computation tasks of each agent~$n$ at each time index~$i$. The gain matrices, $B_i$ and $K_i$, in \eqref{eqn:CIKF_yf}-\eqref{eqn:CIKF_xp} are deterministic and can be pre-computed and saved at each agent. We discuss the details of the design of these optimal gain matrices in Section~\ref{sec:gain}. With these optimal gain matrices, the Kalman type \textit{Consensus$+$Innovations} filter and prediction updates \eqref{eqn:CIKF_yf}-\eqref{eqn:CIKF_xp} provide the minimized MSE distributed estimates of the dynamic states, and hence we term our solution as \textit{Consensus$+$Innovations} Kalman Filter (CIKF).

%
\subsection{CIKF: Assumptions}
\label{subsec:CIKF_assm}
The \textit{Consensus$+$Innovations} Kalman Filter (CIKF) achieves convergence given the following assumptions:
\begin{assumption}[Global detectability]
	\label{assm:detectability}
	The dynamic state equation~\eqref{eqn:sys} and the observations model~\eqref{eqn:obs_global} are globally detectable, i.e., the pair $(A, H)$ is detectable.
\end{assumption}
\begin{assumption}[Connectedness]
	\label{assm:connectivity}
	The agent communication network is connected, i.e., the algebraic connectivity $\lambda_2(L)$ of the Laplacian matrix $L$ of the graph $\mathcal{G}$ is strictly positive.
\end{assumption}

By Assumption~\ref{assm:detectability}, the state-observation model~\eqref{eqn:sys}-\eqref{eqn:obs_global} is globally detectable but not necessarily locally detectable, i.e., $(A, H_n), \forall n, $ are not necessarily detectable. Note that these two are minimal assumptions. Assumption~\ref{assm:detectability} is mandatory even for a centralized system, and Assumption~\ref{assm:connectivity} is required for consensus algorithms to converge. Further, note that in this paper we do not consider distributed observability (invertibility of $G$) of the model setup, which is the strong and restrictive assumption taken in~\cite{das2015TSP},~\cite{kar2012distributed}, \cite{kar2011convergence} and~\cite{shahrampour2016distributed} and similar to weak detectability presented in \cite{kar2011gossip}.

%
\subsection{CIKF: Update algorithm}
\label{subsec:algorithm}
In this subsection, we present the step-by-step tasks executed by each agent~$n$ in the cyber layer to implement the \textit{Consensus$+$Innovations} Kalman Filter (CIKF) and thereby obtain the unbiased minimized MSE distributed estimates of the dynamic state~$\bx_i$. Each agent~$n$ runs Algorithm~\ref{alg:algorithm} locally.
\begin{algorithm}[h]
	\caption{\textit{Consensus$+$Innovations} Kalman Filter}
	\label{alg:algorithm}
	\begin{algorithmic}
		\STATE {\bfseries Input:} Model parameters $A$, $V$, $H$, $R$, $G$, $L$, $\overline{\bx}_0$, $\Sigma_0$.
		\vskip2pt
		\STATE {\bfseries Initialize:} $\widehat{\bx}^n_{0|-1} = \overline{\bx}_0$, $\widehat{\by}^n_{0|-1} = G\overline{\bx}_0$.
		\vskip2pt
		\STATE {\bfseries Pre-compute:} Gain matrices $B_i$ and $K_i$ using Algorithm~\ref{alg:algorithm2}.
		\vskip2pt
		\WHILE{$i \geq 0$}
		\vskip2pt
		\STATE $\!\!\!\!${\textit{Communications:}}
		\vskip2pt
		\STATE Broadcast $\ypp^n$ to all neighbors $l \in \Omega_n$.\\
		\STATE Receive $\{\ypp^l\}_{l \in \Omega_n}$ from neighbors.
		\vskip2pt
		\STATE $\!\!\!\!${\textit{Observation:}}
		\vskip2pt
		\STATE Make measurement $\bz^n_i$ of the state $\bx_i$.\\
		\STATE Transform $\bz^n_i$ in pseudo-observation $\widetilde{\bz}^n_i$ using \eqref{eqn:p_obs_H}.
		\vskip2pt
		\STATE $\!\!\!\!${\textit{Filter updates:}}
		\vskip2pt
		\STATE Compute the estimate $\yf^n$ of $\by_i$ using \eqref{eqn:CIKF_yf}.\\
		\STATE Compute the estimate $\xf^n$ of the state $\bx_i$ using \eqref{eqn:CIKF_xf}.
		\vskip2pt
		\STATE $\!\!\!\!${\textit{Prediction updates:}}
		\vskip2pt
		\STATE Predict the estimate $\yp^n$ of $\by_{i+1}$ using \eqref{eqn:CIKF_yp}. \\
		\STATE Predict the estimate $\xp^n$ of the state $\bx_{i+1}$ using \eqref{eqn:CIKF_xp}. \\
		\vskip2pt
		\ENDWHILE
	\end{algorithmic}
\end{algorithm}

Later in Section~\ref{sec:simulation}, we analyze and compare the performance of CIKF with that of the distributed information Kalman filter (DKF) \cite{das2015TSP} and of the centralized Kalman filter (CKF). The centralized filter collects measurements from all the agents in the cyber layer. The DIKF performs distributed dynamic averaging of the pseudo-observations. The pseudo-state can be perceived as a noise-reduced version of the global average of all the pseudo-observations or pseudo-innovations. The distributed estimates of the pseudo-states reduce the noise of the innovations in the filtering step of field estimation and hence further improve the MSE performance of the distributed estimator. Before going into the numerical evaluation, we do theoretical error analysis of CIKF, derive the conditions for convergence guarantees, and design the optimal consensus and innovations gains in the following sections.

%
\section{Error Analysis}
\label{sec:error}
We analyze the MSE performance of the CIKF and derive its error covariance matrices. First, we define the different error processes and determine their dynamics. Denote the filtering error processes~$\ef^n$ and~$\epsf^n$ of the pseudo-state and of the state at agent~$n$ by
\begin{align}
\label{eqn:efn}
\ef^n &= \by_i - \yf^n, \\
\label{eqn:epsfn}
\epsf^n &= \bx_i - \xf^n.
\end{align}
Similarly, represent the prediction error processes~$\ep^n$ and~$\epsp^n$ of the pseudo-state and the state at agent~$n$ by
\begin{align}
\label{eqn:epn}
\ep^n &= \by_{i+1} - \yp^n, \\
\label{eqn:epspn}
\epsp^n &= \bx_{i+1} - \xp^n.
\end{align}
We establish that the CIKF provides unbiased estimates of the state and pseudo-state in the following lemma whose proof is sketched in Appendix~\ref{app:proof_error}.
\begin{lemma}
	\label{lemma:error_unbiased}
	The distributed filter and prediction estimates, $\yf^n$, $\xf^n$, $\yp^n$ and $\xp^n$, of the pseudo-state and state are unbiased, i.e., error processes, $\ef^n$, $\ep^n$, $\epsf^n$, and $\epsp^n$ are zero-mean at all agents~$n$:
	\begin{align}
	\label{eqn:error_unbiased}
		\mathbb{E}[\ef^n] = 0, \;
		\mathbb{E}[\ep^n] = 0, \;
		\mathbb{E}[\epsf^n] = 0, \;
		\mathbb{E}[\epsp^n] = 0.
	\end{align}
\end{lemma}
Each agent exchanges their estimates with their neighbors, hence their error processes are correlated. Since the error processes of the agents are coupled, we stack them in a error vector and analyze all errors together.
\begin{small}
\begin{align*}
\xf &\!=\! \begin{bmatrix} \xf^1 \\ \vdots \\ \xf^N \end{bmatrix} \!,
\xp \!=\! \begin{bmatrix} \xp^1 \\ \vdots \\ \xp^N \end{bmatrix}\!,
\yf \!=\! \begin{bmatrix} \yf^1 \\ \vdots \\ \yf^N \end{bmatrix}\!,
\yp \!=\! \begin{bmatrix} \yp^1 \\ \vdots \\ \yp^N \end{bmatrix}\!, \\
\epsf &\!=\! \begin{bmatrix} \epsf^1 \\ \vdots \\ \epsf^N \end{bmatrix}\!,
\epsp \!=\! \begin{bmatrix} \epsp^1 \\ \vdots \\ \epsp^N \end{bmatrix}\!,
\ef \!=\! \begin{bmatrix} \ef^1 \\ \vdots \\ \ef^N \end{bmatrix}\!,
\ep \!=\! \begin{bmatrix} \ep^1 \\ \vdots \\ \ep^N \end{bmatrix}\!.
\end{align*}
\end{small}
We summarize the dynamics of the error processes in the following lemma whose proof is in Appendix~\ref{app:proof_unbiased}.
\begin{lemma}
	\label{lemma:error_dynamics}
	The error processes, $\ef$, $\ep$, $\epsf$, and $\epsf$ are Gaussian and their dynamics are:
	\begin{small}
	\begin{align}
	\label{eqn:error_dynamics_ef}
	&\!\!\ef \!=\! \left( I_{\!M\!N} \!-\! B^{\mathcal{C}}_i \!-\! B^{\mathcal{I}}_i \widetilde{D}_{H}\right)\! \epp \!-\! B^{\mathcal{I}}_i \check{D}_{H} \epspp \!-\! B^{\mathcal{I}}_i D_{H}^T R^{-1} r_i , \\
	\label{eqn:error_dynamics_epsf}
	&\epsf = \left( I_{MN} - K_i \left( I_N \otimes G \right) \right) \epspp + K_i \ef, \\
	\label{eqn:error_dynamics_ep}
	&\ep = \left(I_{\!N} \!\otimes\! \widetilde{A}\right)\! \ef + \!\left( I_{\!N} \!\otimes\! \check{A} \right)\! \epsf + 1_{\!N} \!\otimes\! \left(G\bv_i\right), \\
	\label{eqn:error_dynamics_epsp}
	&\epsp = \left( I_N \otimes A\right) \epsf + 1_N \otimes \bv_i,
	\end{align}
	\end{small}
	where, $B^{\mathcal{C}}_i$ is the consensus gain matrix and $B^{\mathcal{I}}_i$, $K_i$ are the innovations gain matrices for the pseudo-state and state estimation, respectively. The block diagonal matrices are $\widetilde{D}_{H} = \text{blockdiag}\{\widetilde{H}_1, \cdots, \widetilde{H}_N \}$ and $\check{D}_{H} = \text{blockdiag}\{\check{H}_1, \cdots, \check{H}_N \}$.
\end{lemma}
The symbol~$\otimes$ denotes the Kronecker matrix product. Lemma~\ref{lemma:error_unbiased} established that the error processes \eqref{eqn:error_dynamics_ef}-\eqref{eqn:error_dynamics_epsp} are unbiased. It then follows that the filter and prediction error covariances of the pseudo-state and state are simply:
\begin{align}
\label{eqn:Pf}
P_{i|i} &= \mathbb{E}\left[\ef \ef^T \right]  \\
\label{eqn:Pp}
P_{i+1|i} &= \mathbb{E}\left[\ep \ep^T \right] \\
\label{eqn:Sigf}
\Sigma_{i|i} &= \mathbb{E}\left[\epsf \epsf^T \right] \\
\label{eqn:Sigp}
\Sigma_{i+1|i} &= \mathbb{E}\left[\epsp \epsp^T \right]
\end{align}
Note that the state estimates~$\xf$,~$\xp$ depend on the pseudo-state estimates~$\yf$,~$\yp$. Hence the error process \eqref{eqn:error_dynamics_ef}-\eqref{eqn:error_dynamics_epsp} are not uncorrelated. The filter and prediction cross-covariances are:
\begin{align}
\label{eqn:Pif}
\Pi_{i|i} &= \mathbb{E}\left[\epsf \ef^T \right] \\
\label{eqn:Pip}
\Pi_{i+1|i} &= \mathbb{E}\left[\epsp \ep^T \right] \\
\label{eqn:Gamma}
\Gamma_{i} &= \mathbb{E}\left[\epspp \ef^T \right]
\end{align}
In the following theorem, we define and derive the evolution of the state, pseudo-state, and cross error covariances.

	\begin{figure*}
		\small
		\setcounter{equation}{40}				
		\begin{small}
			\begin{align}
			\label{eqn:Gamma_dynamics}
			& \Gamma_{i} = \Pi_{i|i-1} \left( I_{MN} - B^{\mathcal{C}}_i - B^{\mathcal{I}}_i \widetilde{D}_{H}\right)^T - \Sigma_{i|i-1}\check{D}_{H}^T B^{\mathcal{I}^T}_i \\
			\label{eqn:Pf_dynamics}
			& P_{i|i} = \left( I_{\!M\!N} \!-\! B^{\mathcal{C}}_i \!-\! B^{\mathcal{I}}_i \widetilde{D}_{H}\right) P_{i|i-1} \left( I_{MN} \!-\! B^{\mathcal{C}}_i \!-\! B^{\mathcal{I}}_i \widetilde{D}_{H}\right)^T \!+\! B^{\mathcal{I}}_i \overline{D}_H B^{\mathcal{I}^T}_i \!-\! \left( I_{\!M\!N} \!-\! B^{\mathcal{C}}_i \!-\! B^{\mathcal{I}}_i \widetilde{D}_{H}\right) \Pi_{i|i-1}^T \check{D}_{H} B^{\mathcal{I}^T}_i \!-\! B^{\mathcal{I}}_i \check{D}_{H} \Gamma_i \\
			\label{eqn:Sigf_dynamics}
			&\! \Sigma_{i|i} \!=\! \left( I_{\!_{M\!N}}  \!\!-\! K_i \! \left( I_{\!N} \!\otimes\! G \right) \right) \! \Sigma_{i|i-1} \! \left( I_{\!_{M\!N}} \!\!-\! K_i \left( I_{\!N} \!\otimes\! G \right) \right) \!+\! K_i P_{i|i} K^T_i \!+\! \left( I_{\!_{M\!N}}  \!\!-\! K_i \! \left( I_{\!N} \!\otimes\! G \right) \right) \! \Pi_{i|i-1} \! K^T_i \!+\! K_i \Pi_{i|i-1}^T \left( I_{\!_{M\!N}} \!\!-\! K_i \! \left( I_{\!N} \!\otimes\! G \right) \right)^T \\
			\label{eqn:Pif_dynamics}
			&\Pi_{i|i} = \left( I_{\!_{M\!N}} \!-\! K_i \! \left( I_{\!N} \!\otimes\! G \right) \right) \Gamma_i + K_i P^T_{i|i}  \\
			\label{eqn:Pp_dynamics}
			& P_{i+1|i} \!=\! \left( I_{\!N} \!\otimes\! \widetilde{A} \right) P_{i|i} \left( I_{\!N} \!\otimes\! \widetilde{A}^T \right) \!+\! \left( I_{\!N} \!\otimes\! \check{A} \! \right) \Sigma_{i|i} \left( I_{\!N} \!\otimes\! \check{A}^T \! \right) \!+\! J\!\otimes\! \left(GVG\right) \!+\! \left( I_{\!N} \!\otimes\! \check{A} \! \right) \Pi_{i|i} \left( I_{\!N} \!\otimes\! \widetilde{A}^T \right) \!+\! \left( I_{\!N} \!\otimes\! \widetilde{A} \right) \Pi_{i|i}^T \left( I_{\!N} \!\otimes\! \check{A}^T \! \right)  \\
			\label{eqn:Sigp_dynamics}
			& \Sigma_{i+1|i} = \left( I_{\!N} \!\otimes\! A \right) \Sigma_{i|i} \left( I_{\!N} \!\otimes\! A^T \right) + J \!\otimes\! V	\\
			\label{eqn:Pip_dynamics}
			& \Pi_{i+1|i} = \left( I_{\!N} \!\otimes\! A \right) \Pi_{i|i} \left( I_{\!N} \!\otimes\! \widetilde{A}^T \right) + \left( I_{\!N} \!\otimes\! A \right) \Sigma_{i|i} \left( I_{\!N} \!\otimes\! \check{A}^T \! \right) + J\!\otimes\! \left(VG\right)
			\end{align}	
		\end{small}
		
		\setcounter{equation}{47}
		\vskip-10pt
		\vspace*{-4pt}
		\hrulefill
	\end{figure*}	
%
\begin{theorem}
	\label{thm:covariance}
	The filter error covariances, $P_{i|i}$, $\Sigma_{i|i}$, $\Pi_{i|i}$, and the predictor error covariances, $P_{i+1|i}$, $\Sigma_{i+1|i}$, $\Pi_{i+1|i}$, follow Lyapunov-type iterations~\eqref{eqn:Pf_dynamics}-\eqref{eqn:Pip_dynamics}, 	
	where, $J = (\bm{1}_N \bm{1}_N^T)\otimes I_{\!M}$ and the initial conditions are $\Sigma_{0|-1} = J\!\otimes\!\Sigma_0, P_{0|-1} = J\!\otimes\!\left(G\Sigma_0 G\right), \Pi_{0|-1} = J\!\otimes\!\left(\Sigma_0 G\right)$.
\end{theorem}

The proof of the theorem is in Appendix~\ref{app:proof_covariance}. The iterations~\eqref{eqn:Sigf_dynamics} and~\eqref{eqn:Sigp_dynamics} combined together constitute the distributed version of the discrete algebraic Riccati equation. The MSE of the proposed CIKF is the trace of the error covariance,~$\Sigma_{i+1|i}$ in~\eqref{eqn:Sigp_dynamics}. The optimal design of the gain matrices,~$B_i$ and~$K_i$, such that the CIKF yields minimized MSE estimates, is discussed in Section~\ref{sec:gain}. Before that in Section~\ref{sec:capacity} we derive the conditions under which the CIKF converges, in other words, the MSE given by the trace of~$\Sigma_{i+1|i}$ is bounded.

%
\section{Tracking Capacity}
\label{sec:capacity}
The convergence properties of the CIKF is determined by the dynamics of the pseudo-state and state error processes,~$\ep$ and~$\epsp$. If the error dynamics are asymptotically stable, then the error processes have asymptotically bounded error covariances that in turn guarantee the convergence of the CIKF. Note that if the dynamics of the prediction error processes,~$\ep$,~$\epsp$ are asymptotically stable, then the dynamics of the filter error processes,~$\ef$,~$\epsf$ are also asymptotically stable. That is why we study the dynamics of only one of the error processes and in this paper we consider the prediction error processes.
%
\subsection{Asymptotic stability of error processes}
\label{subsec:stability}
To analyze the stability of the error processes, we first write the evolution of the prediction error processes, combining~\eqref{eqn:error_dynamics_ef}-\eqref{eqn:error_dynamics_epsp},
\begin{align}
\label{eqn:ep_stability}
\ep &= \underbrace{ \left(I_{\!N} \!\otimes\! \widetilde{A}\right)\! \left( I_{\!M\!N} \!-\! B^{\mathcal{C}}_i \!-\! B^{\mathcal{I}}_i \widetilde{D}_{H}\right) }_{\widetilde{F}} \epp  + \widetilde{\bm \phi}_i, \\
\label{eqn:epsp_stability}
\epsp &= \underbrace{ \left( I_{\!N} \!\otimes\! A\right)\! \left( I_{\!M\!N} \!-\!\! K_i \left( I_{\!N} \!\otimes\! G \right) \right) }_{F} \! \epspp \!+\! {\bm \phi}_i ,
\end{align}
where, the noise processes~$\widetilde{\bm \phi}_i$ and~${\bm \phi}_i$ are
\begin{small}
\begin{align}
\widetilde{\bm \phi}_i &= \!\left( I_{\!N} \!\otimes\! \check{A}\right)\! \epsf - \left(I_{\!N} \!\otimes\! \widetilde{A}\right)\!B^{\mathcal{I}}_i \check{D}_{H} \epspp + 1_{\!N} \!\otimes\! \left(G\bv_i\right) \nonumber \\
\label{eqn:til_phi}
& \qquad  - \left(I_{\!N} \!\otimes\! \widetilde{A}\right)\! B^{\mathcal{I}}_i D_{H}^T R^{-1} r_i, \\
\label{eqn:phi}
{\bm \phi}_i &= \left( I_{\!N} \!\otimes\! A\right)\! K_i \ef + 1_N \otimes \bv_i.
\end{align}
\end{small}
The statistical properties of the noises,~$\widetilde{\bm \phi}_i$ and~${\bm \phi}_i$, of the error processes,~$\ep$ and~$\epsp$, are stated in the following Lemma, and the proof is included in Appendix~\ref{app:proof_noises}.
\begin{lemma}
	\label{lemma:noises}
	The noise sequences~$\widetilde{\bm \phi}_i$ and~${\bm \phi}_i$ are zero-mean Gaussian that follow~$\widetilde{\bm \phi}_i \in \mathcal{N} \left( {\bm 0}, \widetilde{\Phi}_i \right)$ and~${\bm \phi}_i \in \mathcal{N} \left( {\bm 0}, \Phi_i \right)$.
\end{lemma}
The dynamics of the error processes are characterized by \eqref{eqn:ep_stability}-\eqref{eqn:epsp_stability} and Lemma~\ref{lemma:noises}. Let $\rho(.)$ and $\|.\|_2$ denote the spectral radius and the spectral norm of a matrix, respectively. The error processes are asymptotically stable if and only if the spectral radii of~$\widetilde{F}$,~$F$ are less than one, i.e.,
\begin{align}
\label{eqn:stability}
\rho\left(\widetilde{F}\right) < 1, \quad \rho\left(F\right) < 1
\end{align}
and the noise covariances,~$\widetilde{\Phi}_i$,~$\Phi_i$ are bounded, i.e., $\|\widetilde{\Phi}_i\|_2 < \infty$,~$\|\Phi_i\|_2 < \infty, \; \forall i$. Now if~\eqref{eqn:stability} holds, then the prediction error covariances~$P_{i+1|i}$,~$\Sigma_{i+1|i}$ are bounded; this ensures the filter error covariances~$P_{i|i}$,~$\Sigma_{i|i}$ are also bounded. Further, the model noise covariances $V$ and $R$ are bounded. Then, by~\eqref{eqn:til_Phi_dynamics}-\eqref{eqn:Phi_dynamics}, the noise covariances~$\widetilde{\Phi}_i$ and~${\Phi}_i$ are bounded if the spectral radii are less than one. Thus,~\eqref{eqn:stability} are the necessary and sufficient conditions for the convergence of the CIKF algorithm.
%
\subsection{Tracking capacity for unstable systems}
\label{subsec:tracking_unstable}
The stability of the underlying dynamical system~\eqref{eqn:sys} in the physical layer is determined by the dynamics matrix~$A$. If the system is asymptotically stable, i.e., $\rho(A)<1$, then there always exist gain matrices~$B_i$,~$K_i$ such that~\eqref{eqn:stability} holds true. Hence for stable systems, the CIKF always converges with a bounded MSE solution. In contrast, for an unstable dynamical system~\eqref{eqn:sys}, $\rho(A)>1$, it may not always be possible to find gain matrices ~$B_i$,~$K_i$ satisfying~\eqref{eqn:stability} conditions. There exists an upper threshold on the degree of instability of the system dynamics,~$A$, that guarantees the convergence of the proposed CIKF. The threshold, similar to Network Tracking Capacity in~\cite{khan2010connectivity}, is the tracking capacity of the CIKF algorithm, and it depends on the agent communication network and observation models, as summarized in the following theorem.
\begin{theorem}
	\label{thm:capacity}
	The tracking capacity of the CIKF is, C,
	\begin{align}
	\label{eqn:capacity}
	C = \max_{B^{\mathcal{C}}_i, B^{\mathcal{I}}_i} \frac{\lambda_{1}}{\lambda_{m} \begin{Vmatrix} I_{\!M\!N} \!-\! B^{\mathcal{C}}_i \!-\! B^{\mathcal{I}}_i \widetilde{D}_{H} \end{Vmatrix}_2 }
	\end{align}
	where, $B^{\mathcal{C}}_i$ has the same block sparsity pattern as the graph Laplacian~$L$, $B^{\mathcal{I}}_i$~is a block diagonal matrix, and, $\lambda_{1}$~and $\lambda_{m}$~are the minimum and maximum non-zero eigenvalues of~$G$, $0<\lambda_{1} \leq \cdots \leq \lambda_{m}$. If $\| A \|_2 < C$, then there exists~$B_i$,~$K_i$ such that the CIKF~\eqref{eqn:CIKF_yf}-\eqref{eqn:CIKF_xp} converges with bounded MSE.
\end{theorem}
The proof is in Appendix~\ref{app:proof_capacity}. In the above theorem, the structural constraints on the gain matrices $B^{\mathcal{C}}_i, B^{\mathcal{I}}_i$ ensure that each agent combines its neighbors' estimates for consensus and its own pseudo-observations for the innovation part of the CIKF. The block sparsity pattern of~$B^{\mathcal{C}}_i$ being similar to that of~$L$ implies that the tracking capacity is dependent on the connectivity of the communication network. Similarly, since~$\widetilde{D}_{H}$ is a block-diagonal matrix containing the observation matrices~$H_n$ we conclude that the tracking capacity is also a function of the observation models.

The tracking capacity increases with the increase in communication graph connectivity and observation density. For instance, the tracking capacity is infinity if all agents are connected with everyone else (complete graph) or all the agents observe the entire dynamical system (local observability). Given the tracking capacity is satisfied for the system, observation, and communication models \eqref{eqn:sys}-\eqref{eqn:comm}, the question remains how to design the gain matrices~$B_i$ and~$K_i$ to minimize the MSE of the CIKF, which we discuss in the following section.

%
\section{Optimal Gain Design}
\label{sec:gain}
The asymptotic stability of the error dynamics guarantees convergence of CIKF and bounded MSE, but here we discuss how to design the $B_i$ and $K_i$ such that the MSE is not only bounded but also minimum.

%
\subsection{New uncorrelated information}
\label{subsec:uncorrelated}
In CIKF Algorithm~\ref{alg:algorithm}, at any time~$i$ each agent~$n$ makes pseudo-observation $\widetilde{\bz}^n_i$ of the state and receives prior estimates $\{\ypp^l\}_{l \in \Omega_n}$ from its neighbors. The CIKF algorithm employs this new information to compute the distributed filter estimates of the pseudo-state and state. Denote the new information for the pseudo-state and state filtering by~$\widetilde{\bm \theta}^n_i$ and~${\bm \theta}^n_i$, respectively,
\begin{align}
\label{eqn:new_info}
\widetilde{\bm \theta}^n_i = \begin{bmatrix}\ypp^{l_1} \\ \vdots \\  \ypp^{l_{d_n}} \\ \widetilde{\bz}^n_i \end{bmatrix}, \quad
{\bm \theta}^n_i = \yf^n
\end{align}
where, $\{ l_1, \cdots, l_{d_n}\} = \Omega_n$ and $d_n = |\Omega_n|$ is the degree of agent~$n$. Note the new information,~$\widetilde{\bm \theta}^n_i$ and~${\bm \theta}^n_i$, are Gaussian since they are linear combinations of Gaussian sequences. However,~$\widetilde{\bm \theta}^n_i$ and~${\bm \theta}^n_i$ are correlated with the previous estimates~$\ypp^n$ and~$\xpp^n$. So, we transform them into uncorrelated new information and then combine the uncorrelated information with the previous estimates~$\ypp^n$ and~$\xpp^n$ to compute the current filtered estimates.
\begin{lemma}
	\label{lemma:uncorrelation}
	The new uncorrelated information~$\widetilde{\bm \nu}^n_i$ and~${\bm \nu}^n_i$ for filtering update at agent~n are,
	\begin{small}
		\begin{align}
		\label{eqn:nu1_tilde}
		\widetilde{\bm \nu}^n_i &= \widetilde{\bm \theta}^n_i - \overline{\widetilde{\bm \theta}}^n_i, \qquad \overline{\widetilde{\bm \theta}}^n_i = \mathbb{E} \begin{bmatrix} \widetilde{\bm \theta}^n_i | \widetilde{\bm z}^n_{i-1}, \{ \widehat{\bm y}^l_{i-1|i-2} \}_{l \in \Omega_n} \end{bmatrix} \\
		\label{eqn:nu1}
		{\bm \nu}^n_i &= {\bm \theta}^n_i - \overline{\bm \theta}^n_i, \qquad \overline{\bm \theta}^n_i = \mathbb{E} \begin{bmatrix} {\bm \theta}^n_i | \widehat{\bm y}^n_{i-1|i-1} \end{bmatrix}
		\end{align}
	\end{small}
	that expands to
	\begin{small}
	\begin{align}
	\label{eqn:nu}
		\widetilde{\bm \nu}^n_i \!=\! \begin{bmatrix}\ypp^{l_1} - \ypp^n \\ \vdots \\  \ypp^{l_{d_n}}-\ypp^n \\ \widetilde{\bz}^n_i \!\!-\! \widetilde{H}_{\!n} \ypp^n \!-\! \check{H}_{\!n} \widetilde{I} \xpp^n \end{bmatrix}\!\!\!, \;\;
		{\bm \nu}^n_i \!=\!  \yf^n \!-\! G\xpp^n.
	\end{align}
	\end{small}
	The uncorrelated sequences~$\widetilde{\bm \nu}^n_i$ and~${\bm \nu}^n_i$ are zero-mean Gaussian random vectors. Hence~$\widetilde{\bm \nu}^n_i$ and~${\bm \nu}^n_i$ are independent sequences.
\end{lemma}
The proof is in Appendix~\ref{app:proof_uncorrelation}. We write the CIKF filter updates~\eqref{eqn:CIKF_yf} and~\eqref{eqn:CIKF_xf} in terms of the new uncorrelated information~$\widetilde{\bm \nu}^n_i$ and~${\bm \nu}^n_i$ from~\eqref{eqn:nu},
\begin{align}	
	\label{eqn:CIKF_yf_nu}	
	\yf^n &= \ypp^n + \widehat{B}^n_i \widetilde{\bm \nu}^n_i \\
	\label{eqn:CIKF_xf_nu}	
	\xf^n &= \xpp^n + K^n_i {\bm \nu}^n_i
\end{align}
where, $\widehat{B}^n_i$ are the building blocks of the pseudo-state gain matrix~$B_i$.
%

\subsection{Consensus and innovation gains}
\label{subsec:gains}
Here, we present the methods to: (a) design the matrices~$\widehat{B}^n_i$ and~$K^n_i$; and (b) obtain the optimal gains~$B_i$ and~$K_i$ from them. These optimal gains provide the distributed minimized MSE estimates of the field. At agent~$n$, we define the matrix~$\widehat{B}^n_i$ as
\begin{align}
\label{eqn:Bn_temp}
\widehat{B}^n_i = \begin{bmatrix} B^{n l_1}_i, \cdots, B^{n l_{d_n}}_i, B^{n n}_i \end{bmatrix}
\end{align}
where, $\{ l_1, \cdots, l_{d_n}\} = \Omega_n$. The gain matrix~$B_i$ is a linear combination of $B^{\mathcal{C}}_i$ and $B^{\mathcal{I}}_i$, where $B^{\mathcal{C}}_i$ has the same block structure as the graph Laplacian~$L$ and~$B^{\mathcal{I}}_i$ is a block diagonal. The $(n,l)^{\text{th}}$ blocks of the $n^{\text {th}}$ row of $B^{\mathcal{C}}_i$ are:
\begin{align}
\label{eqn:BC}
\left[B^{\mathcal{C}}_i\right]_{nl} = \left\{
\begin{array}{ll}
- B^{n l}_i,  & \mbox{if } l \in \Omega_n  \\
\sum_{j=1}^{d_n}B^{n l_j}_i, & \mbox{if } l = n \\
{\bm 0}, & \text{otherwise.}  \\
\end{array} \right.
\end{align}
The $\{n,n\}^{\text{th}}$ blocks of the diagonal block matrices $B^{\mathcal{I}}_i$ and $K_i$ are:
\begin{align}
\label{eqn:BI}
\left[B^{\mathcal{I}}_i\right]_{nn} &= B^{n n}_i, \\
\label{eqn:K_design}
\left[K_i\right]_{nn} &= K^{n}_i,
\end{align}
Hence, once we design the matrices~$\widehat{B}^n_i$ and~$K^{n}_i$, it will provide the optimal gain matrices~$B^{\mathcal{C}}_i, B^{\mathcal{I}}_i$, and $K_i$.
\begin{theorem}
	\label{thm:gain}
	The optimal gains for the CIKF algorithm are
	\begin{align*}
		\widehat{B}^n_i &= \Sigma_{\by_i \widetilde{\bm \nu}^n_i} \left(\Sigma_{\widetilde{\bm \nu}^n_i}\right)^{-1}, \\
		K^n_i &= \Sigma_{\bx_i {\bm \nu}^n_i} \left(\Sigma_{{\bm \nu}^n_i}\right)^{-1}
	\end{align*}
	where, $\Sigma_{\widetilde{\bm \nu}^n_i}$, $\Sigma_{{\bm \nu}^n_i}$ are the covariances of the new uncorrelated information~$\widetilde{\bm \nu}^n_i$ and~${\bm \nu}^n_i$; and, $\Sigma_{\by_i \widetilde{\bm \nu}^n_i}$, $\Sigma_{\bx_i {\bm \nu}^n_i}$ are cross-covariances between $\by_i$, $\widetilde{\bm \nu}^n_i$ and $\bx_i$, ${\bm \nu}^n_i$, respectively. These covariance and cross-covariance matrices are related to the error covariance matrices, $P_{i|i-1}, P_{i|i}, \Sigma_{i|i-1}, \Pi_{i|i-1}, \Gamma_i$, by the following functions:
	\begin{small}
	\begin{align*}
	& \!\Sigma_{\by_i \widetilde{\bm \nu}^n_i} \!=\!  \left[
	\begin{array}{l;{2pt/2pt}l;{1pt/1pt}l;{2pt/2pt}l}
	\!\!\!\! P_{i|i\!-\!1}^{nn} \!-\! P_{i|i\!-\!1}^{nl_1} & \cdots & P_{i|i\!-\!1}^{nn} \!-\! P_{i|i\!-\!1}^{nl_{d_n}} & P_{i|i\!-\!1}^{nn} \! \widetilde{H}^T_n \!+\! \Pi_{i|i\!-\!1}^{nn^T} \! \check{H}^T_n \!\!\!\!
	\end{array}
	\right] \\
	&\! \Sigma_{\bx_i {\bm \nu}^n_i} = \Sigma_{i|i\!-\!1}^{nn} G - \Gamma^{nn}_i \\
	&\begin{bmatrix} \Sigma_{\widetilde{\bm \nu}^n_i} \end{bmatrix}_{qs} = \\
	&\left\{
		\begin{array}{ll}
		P_{i|i\!-\!1}^{nn} \!-\! P_{i|i\!-\!1}^{nl_s} \!-\! P_{i|i\!-\!1}^{l_q n} \!+\! P_{i|i\!-\!1}^{l_q l_s},  & \mbox{if } q \leq s \leq d_n \\
		\!\left( P_{i|i\!-\!1}^{nn} \!-\! P_{i|i\!-\!1}^{l_q n} \right)\!\widetilde{H}_n^T \!+\! \left( \Pi_{i|i\!-\!1}^{nn} \!-\! \Pi_{i|i\!-\!1}^{l_q n} \right)^T\!\!\check{H}_n^T, & \mbox{if } q < s = d_n\!\!+\!1 \\
		\widetilde{H}_n P_{i|i\!-\!1}^{nn}\widetilde{H}_n^T + \widetilde{H}_n \Pi_{i|i\!-\!1}^{nn^T}\check{H}_n^T  + \check{H}_n\Pi_{i|i\!-\!1}^{nn}\widetilde{H}_n^T & \\
		\qquad \qquad + \check{H}_n\Sigma_{i|i\!-\!1}^{nn}\check{H}^T_n + \overline{H}_n, & \mbox{if } q = s = d_n\!\!+\!1 \\ %
		\begin{bmatrix} \Sigma_{\widetilde{\bm \nu}^n_i} \end{bmatrix}_{sq}^T, & \mbox{if } q > s \\
		\end{array} \right. \\
	&\begin{bmatrix} \Sigma_{{\bm \nu}^n_i} \end{bmatrix}_{qs}  = G\Sigma_{i|i\!-\!1}^{nn}G - G \Gamma^{nn}_i - \Gamma^{nn^T}_iG + P_{i|i}^{nn}
	\end{align*}	
	\end{small}
	where, $\left[\Sigma_{\widetilde{\bm \nu}^n_i}\right]_{qs}$ denotes the $\{q,s\}^{\text{th}}$ block of the $(d_n\!\!+\!1)\times(d_n\!\!+\!1)$ block matrix~$\Sigma_{\widetilde{\bm \nu}^n_i}$.
\end{theorem}
The proof is in Appendix~\ref{app:proof_gain}. By the Gauss-Markov theorem, the CIKF algorithm, along with this design of the consensus and innovation gain matrices, as stated in Theorem~\ref{thm:gain}, results in the minimized MSE distributed estimates of the dynamic random field~$\bx_i$. The gain matrices are deterministic. Hence they can be precomputed offline and saved for online implementation. In Algorithm~\ref{alg:algorithm2}, we state the steps that each agent~$n$ runs to compute the optimal gain matrices.

\begin{algorithm}[h]
	\caption{Gain Design of CIKF}
	\label{alg:algorithm2}
	\begin{algorithmic}
		\STATE {\bfseries Input:} Model parameters $A$, $V$, $H$, $R$, $G$, $L$, $\Sigma_0$.
		\vskip2pt
		\STATE {\bfseries Initialize:} $\Sigma_{0|-1} = J \!\otimes\! \Sigma_0, \; P_{0|-1} = J \!\otimes\! \left(G\Sigma_0 G\right), \; \Pi_{0|-1} = J \!\otimes\! \left(\Sigma_0 G\right)$.
		\vskip2pt
		\WHILE{$i \geq 0$}
		\vskip2pt
		\STATE $\!\!\!\!${\textit{Optimal gains:}}
		\vskip2pt
		\STATE Compute $\widehat{B}^n_i$ and~$K^n_i$ using Theorem~\ref{thm:gain}.\\
		\STATE Using \eqref{eqn:BC}-\eqref{eqn:K_design}, obtain $B^{\mathcal{C}}_i$, $B^{\mathcal{I}}_i$ and $K_i$ from $\widehat{B}^n_i$ and~$K^n_i$.
		\vskip2pt
		\STATE $\!\!\!\!${\textit{Prediction error covariance updates:}}
		\vskip2pt
		\STATE Update $P_{i+1|i}, \Sigma_{i+1|i}, \Pi_{i+1|i}$ using~\eqref{eqn:Pf_dynamics}-\eqref{eqn:Pip_dynamics}.
		\vskip2pt
		\ENDWHILE
	\end{algorithmic}
\end{algorithm}

The offline Algorithm~\ref{alg:algorithm2} along with the online Algorithm~\ref{alg:algorithm} completes our proposed distributed solution to obtain minimized MSE estimates of the dynamic field~$\bx_i$ at each agent in the cyber network. CIKF does not require sparsity of the matrices $A$, $H_n$ and $L$. However, in most distributed applications these matrices are sparse. For example, consider the temperature distribution over the entire United States as a time-varying random field. It is conceivable that the temperature in a location will be directly dependent on the temperature of a few locations leading to a sparse~$A$. Similarly a weather station will make localized measurements of the temperature (at its own location) rather than measuring the temperature all over the US. Similarly the weather stations will exchange their data with the neighboring weather stations, thereby reducing the communication cost and leading to a sparse graph Laplacian matrix. From a computational point of view, if the matrices $A$, $H_n$ and $L$ are sparse, they can be stored as lists, for example, and computation of matrix products with them will involve less multiplications and/or additions operations. Since the computation of $B_i$ and $K_i$ involve these matrices, their sparsity will reduce both the computation and storage costs.		
		
%
\section{Numerical Evaluation}
\label{sec:simulation}
We numerically evaluate the MSE performance of the CIKF and compare it against the centralized Kalman filter (CKF) and the distributed information Kalman filter (DIKF) in~\cite{das2015TSP}. To this objective, we build a time-varying random system, observation and network model that satisfies the Assumptions~\ref{assm:gauss}-\ref{assm:connectivity}. The Algorithms~\ref{alg:algorithm}-\ref{alg:algorithm2} run on these model parameters. First the Algorithm~\ref{alg:algorithm2} computes and save the gain matrices and the error covariance matrices. The traces of the error covariance matrices provide the theoretical MSE trajectory of the CIKF with time. Then, we Monte-Carlo simulate Algorithm~\ref{alg:algorithm} to compute the numerical MSE of the distributed estimators, CIKF and DIKF, and the centralized estimator CKF.

\subsection{Model specifications}
\label{subsec:specs}
Here, we consider a time-varying field,~$\bx_i$, with dimension~$M=50$. The physical layer, consisting of~$M=50$ sites, is monitored by a cyber layer consisting of~$N=50$ agents. Each agent in the cyber layer observes~$M_n=2$ sites of the physical layer. We build the field dynamics matrix~$A$ to be sparse and distributed. The dynamics~$A$ possess the structure of a Lattice graph, where the time evolution of a field variable depends on the neighboring field variables. The structure of the Lattice graph is randomly generated using R library. The degree is considered to be 4 and the non-zero entries are randomly chosen from~$(0,1)$. For illustration, we consider an unstable field dynamics with~$\|A\|_2=1.05$ to test the resilience of the algorithms under unstable conditions. We scale the randomly generated~$A$ such that its spectral norm becomes 1.05. The observation matrices,~$H_n \in \mathbb{R}^{2\times50}, n=1, \cdots, 50$, are sparse~$0-1$ matrices with one non-zero element at each row corresponding to the site of~$\bx_i$ observed by the~$n^{\text{th}}$ agent. The local observations~$\bz^n_i$ are~$2\times1$ random vectors. The mean~$\overline{x}_0$ of the initial state vector is generated at random. The system noise covariance~$V$, the observation noise covariances~$R_n$, and the initial state covariance~$\Sigma_0$ are randomly generated symmetric positive definite matrices. We then scale the covariance matrices such that their covariance becomes: $\|V\|_2=4, \|R_n\|_2=8$, and $\|\Sigma_0\|_2=16$. The agents in the cyber layer communicate among themselves following a randomly generated Erd\H{o}s-R\'enyi graph~$\mathcal{G}$ with~$50$ nodes and~$E=138$ edges. The average degree of each node/agent is approximately $5.5$. The communication network~$\mathcal{G}$ is also sparse.

For Monte-Carlo simulations, we generate the noises,~$\bv_i, \br^n_i$, and the initial condition,~$\bx_0$ as Gaussian sequences, with~$\bv_i \sim \mathcal{N}(\bar 0,V), \;
\br^n_i \sim \mathcal{N}(\bar 0,R_n), \;
\bx_0 \sim \mathcal{N}(\bar{\bx}_0,\Sigma_0)$. The sequences $\{ \{\bv_i\}_i, \{\br_i\}_i, \bx_0\}_{i \geq 0}$ are generated to be uncorrelated. Each agent~$n$ in the cyber layer has access to the system parameters $A, V, H, R, \bar{\bx}_0, \Sigma_0$, and~$\mathcal{G}$. This numerical model satisfies Assumptions~\ref{assm:gauss}-\ref{assm:info}. The pair~$(A,H)$ is detectable and the pairs,~$(A,H_n) \; \forall n$, are not detectable. The agent communication graph~$\mathcal{G}$ is connected with the algebraic connectivity of the Laplacian~$\lambda_2(L) = 0.7 > 0$. Hence the Assumptions~\ref{assm:detectability}-\ref{assm:connectivity} hold true for this numerical system, observation, and network model.

\begin{figure*}
	\centering
	\begin{subfigure}[b]{0.42\textwidth}
		\includegraphics[width=\textwidth]{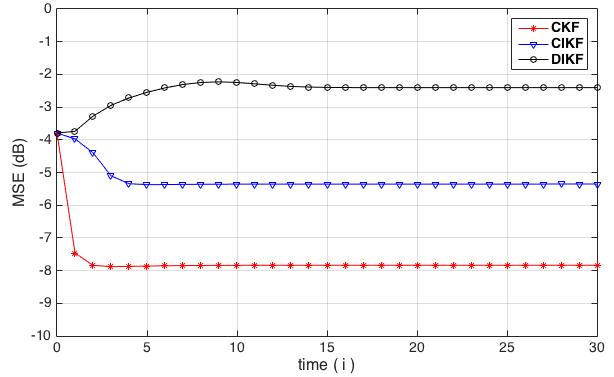}
		\caption{Theoretical MSE}
		\label{fig:mse_th}
	\end{subfigure}%
	~~~~~~~~~
	\begin{subfigure}[b]{0.42\textwidth}
		\includegraphics[width=\textwidth]{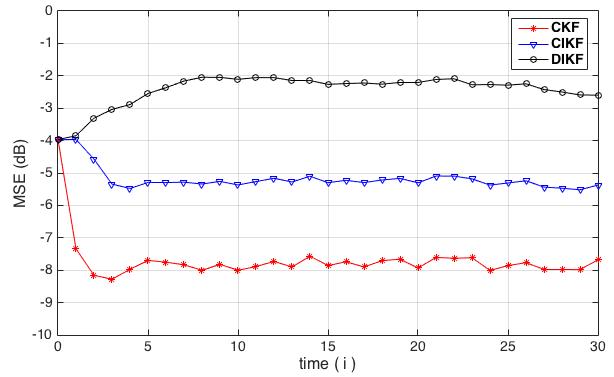}
		\caption{Monte-Carlo MSE}
		\label{fig:mse_sim}
	\end{subfigure}
	\caption{\small Comparison of MSE performance of the proposed CIKF with CKF and DIKF.}
	\vskip5pt
	\hrule
	\vskip-8pt
	\label{fig:sim}
\end{figure*}

\subsection{Optimized gains and theoretical MSE}
\label{subsec:OptGains}
We run Algorithm~\ref{alg:algorithm2} on the numerical model to obtain the gain matrices and the theoretical error covariances of the CIKF. We compute the gain matrices and the error covariances of the centralized Kalman filter (CKF) and of the distributed information Kalman filter (DIKF). In Fig.~\ref{fig:mse_th}, we plot the MSE (in dB), trace of the predictor error covariance matrices~$\Sigma_{i+1|i}$, for each of these cases up to time~$i=30$. The MSE of the optimal CKF is the smallest (recall the CKF, if feasible, would be optimal) and the objective of the distributed estimators is to achieve MSE performance as close as possible to that of the CKF.

From Fig.~\ref{fig:mse_th}, we see that the MSE of the proposed CIKF is~3dB more than the CKF but is~3dB less than the DIKF. From the plot we see that the CIKF converges faster than the DIKF. Hence, the proposed CIKF provides faster convergence and~3dB MSE performance improvement over the DIKF. The performance of the CKF is 3dB better than the CIKF's due to the fact that the CKF has access to the observations of all the sensors at every time steps. In contrast in CIKF, each agent has access to its own observations and the current estimates of its neighbors only; the impact of the observations from the other agents propagate through the network with delay. As the time-varying field~$\bx_i$ is evolving with input noise~$\bv_i$, lack of access to all the observations containing the driving input~$\bv_i$, combined with the network diffusion delay, causes a performance gap between the CKF and the CIKF.

\subsection{Monte-Carlo Simulations}
\label{subsec:MonteCarlo}
We empirically compute the MSE of the distributed estimates given by the CIKF Algorithm~\ref{alg:algorithm}. We implement the algorithms using Matlab in the Microsoft Azure cloud. Given the computation load because of the large system (M = 50) and network (N = 50) models, we run our simulation on Azure DS13 (8 cores, 56 GB memory) virtual machine (VM). The MSE computation for the CIKF, CKF, and DIKF algorithms with $1000$~Monte-Carlo runs require approximately 30 hours in the Azure DS13 VM. Once we obtain the field prediction estimates for the three algorithms, we compute the empirical prediction error covariance matrices~$\widehat{\Sigma}_{i+1|i}$ and then obtain the Monte-Carlo MSE (in dB) from their trace. From the Monte-Carlo MSE plot in Fig~\ref{fig:mse_sim}, we see that the the empirical plots follow closely the theoretical plots in Fig~\ref{fig:mse_th}.

Both the theoretical Fig~\ref{fig:mse_th} and Monte-Carlo simulated Fig~\ref{fig:mse_sim} MSE performance confirms our CIKF analysis in Sections~\ref{sec:error}-\ref{sec:gain}. The novel \textit{Consensus$+$Innovations} Kalman Filter (CIKF) proposed in Section~\ref{sec:solution} along with the optimized gain designs in Section~\ref{sec:gain} provides unbiased distributed estimates with bounded and minimized MSE for the \textit{Consensus$+$Innovations} distributed solution. The CIKF achieves nearly 3dB better performance than the DIKF~\cite{das2015TSP}.

\section{Conclusions}
\label{sec:conclusion}
\noindent {\textbf{Summary:}} In this paper, we propose a \textit{Consensus$+$Innovations} Kalman Filter (CIKF) that obtains unbiased minimized MSE distributed estimates of the pseudo-states and real-time employs them to obtain the unbiased distributed filtering and prediction estimates of the time-varying random state at each agent. The filter update iterations are of the {\textit{Consensus$+$Innovations}} type. Using the Gauss-Markov principle, we designed the optimal gain matrices that yield approximately 3dB improvement over previous available distributed estimators like the DIKF in~\cite{das2015TSP}.

\noindent {\textbf{Contributions:}} The primary contributions of this paper are: (a) design of a filter and corresponding gain matrices to obtain minimized MSE distributed estimates at each agent under minimal assumptions; and (b) a theoretical characterization of the tracking capacity and distributed version of the algebraic Riccati equation. Compared to DIKF, the contributions of this paper are: \\
	$(a)$ The CIKF is a new algorithm for distributed estimation of time-varying random fields which provides 3dB MSE performance over the DIKF and much closer to the performance of the optimal centralized Kalman filter. \\
	$(b)$ The CIKF does not require the strict distributed observability assumption, i.e., invertibility of the matrix $G$. The DIKF require that $G^{-1}$ exists, which demands that all state variables be observed by at least once agent in the network.\\ 
	$(c)$ In CIKF, we design the optimized gain matrices for the dynamic averaging of pseudo-states using the Gauss-Markov Theorem. Hence, the distributed dynamic averaging step of the CIKF has lower MSE than that of the DIKF. \\
	$(d)$ These advantages are obtained by using pseudo-state and this paper demonstrates its usefulness in distributed field estimation.

\appendix
\section{Appendix}
The Appendices prove the proposition, lemmas, and theorems stated in the paper.

\begin{center}
	{\bf Proof of Proposition}
	\vspace{-8pt}
\end{center}

\subsection{Proof of Proposition \ref{prop:state_space}}
\label{app:proof_pstate}
First we derive the dynamics~\eqref{eqn:pstate_dynamics} of the pseudo-state~$\by_i$. Using~\eqref{eqn:pstate} and~\eqref{eqn:sys},
\begin{align*}
\by_{i+1} &= G\bx_{i+1} \\
&= G \left( A\bx_i + \bv_i\right) \\
&= GA \left( G^{\dagger}G + \widetilde{I}\right) \bx_i + G\bv_i, \quad \begin{small} \left[ \text{by \eqref{eqn:p_identity_mat}}, \; I = G^{\dagger}G + \widetilde{I} \right] \end{small} \\
&= GAG^{\dagger}\by_i + G\bv_i + GA\widetilde{I}\bx_i \\
&= \widetilde{A}\by_i + G\bv_i + \check{A}\bx_i, \qquad \quad \left[ \text{by \eqref{eqn:pstate}} \right].
\end{align*}
Now, we derive the observations~\eqref{eqn:pstate_obs} of the pseudo-observations~$\widetilde{\bz}^n_i$. Using~\eqref{eqn:pseudo_obs} and~\eqref{eqn:obs},
\begin{align*}
\widetilde{\bz}^n_i &= H^T_n R_n^{-1} H_n\bx_i + H^T_n R_n^{-1} \br^n_i \\
&= H^T_nR_n^{-1}H_n \left( G^{\dagger}G + \widetilde{I}\right) \bx_i + H^T_n R_n^{-1} \br^n_i \\
&= H^T_nR_n^{-1}H_n G^{\dagger}G \bx_i + H^T_n R_n^{-1} \br^n_i + H^T_nR_n^{-1}H_n \widetilde{I} \bx_i \\
&= \widetilde{H} \by_i + H^T_n R_n^{-1} \br^n_i + \check{H}_n \bx_i. \qquad \qquad \qquad \qed
\end{align*}


\begin{center}
	{\bf Proof of Lemmas}
	\vspace{-8pt}
\end{center}
%
%
\subsection{Proof of Lemma \ref{lemma:error_unbiased}}
\label{app:proof_error}
Consider the filtering error definitions~\eqref{eqn:efn}-\eqref{eqn:epsfn}. We take expectations on both sides,
\begin{align*}
\mathbb{E}\left[\ef^n\right] &= \mathbb{E}\left[\by_i - \yf^n \right] \\
&= \mathbb{E} \left[ \mathbb{E} \left[ \by_i - \yf^n~|~\widetilde{\bz}^n_i, \{ \ypp^l \}_{l \in \Omega_n}  \right] \right] \\
&= \mathbb{E} \left[ \yf^n - \yf^n \right] = 0 \qquad \left[ \text{by} \eqref{eqn:MMSE_yf} \right] \\
\mathbb{E}\left[\epsf^n\right] &= \mathbb{E}\left[\bx_i - \xf^n \right] \\
&= \mathbb{E} \left[ \mathbb{E} \left[ \bx_i - \xf^n~|~\yf^n  \right] \right] \\
&= \mathbb{E} \left[ \xf^n - \xf^n \right] = 0 \qquad \left[ \text{by} \eqref{eqn:epsfn} \right].
\end{align*}
Similarly, taking expectations on prediction errors~\eqref{eqn:epn}-\eqref{eqn:MMSE_xf},
\begin{align*}
\mathbb{E}\left[\ep^n\right] &= \mathbb{E}\left[\by_{i+1} - \yp^n \right] \\
&= \mathbb{E} \left[ \mathbb{E} \left[ \by_{i+1} - \yp^n~|~\widetilde{\bz}^n_i, \{ \ypp^l \}_{l \in \Omega_n}  \right] \right] \\
&= \mathbb{E} \left[ \yp^n - \yp^n \right] = 0 \qquad \left[ \text{by} \eqref{eqn:MMSE_yp} \right] \\
\mathbb{E}\left[\epsp^n\right] &= \mathbb{E}\left[\bx_{i+1} - \xp^n \right] \\
&= \mathbb{E} \left[ \mathbb{E} \left[ \bx_{i+1} - \xp^n~|~\yf^n \right] \right] \\
&= \mathbb{E} \left[ \xp^n - \xp^n \right] = 0 \qquad \left[ \text{by} \eqref{eqn:epspn} \right]. \qquad \qed
\end{align*}


\subsection{Proof of Lemma \ref{lemma:error_dynamics}}
\label{app:proof_unbiased}

We write the pseudo-state filtering update~\eqref{eqn:CIKF_yf} in vector form,
\begin{align*}
\yf \!=\! \ypp & \!-\!  B^{\mathcal{C}}_i \ypp \!+\! B^{\mathcal{I}}_i \! \left( \!\widetilde{\bz}_i \!-\! \left(\widetilde{D}_H \ypp \!+\! \check{D}_H \xpp \right) \right) ,
\end{align*}
where, $\left[B^{\mathcal{C}}_i\right]_{nl} = -B^{n,l}_i, \; \forall n \neq l$, $\left[B^{\mathcal{C}}_i\right]_{nn} = \sum_{l \in \Omega_n}B^{n,l}_i$, $\left[B^{\mathcal{I}}_i\right]_{nn} = B^{n,n}_i$, $\left[B^{\mathcal{I}}_i\right]_{nl} = 0, \; \forall n \neq l$. The block-diagonal matrices: $\widetilde{D}_H = \text{blockdiag}\{\widetilde{H}_1, \cdots, \widetilde{H}_N\}, \; \check{D}_H = \text{blockdiag}\{\check{H}_1, \cdots, \check{H}_N\}$. Note that $B^{\mathcal{C}}_i \left(1_N\otimes\by_i\right) = \bm{0}$. Using this relation and the vector form of~$\yf$, we expand the pseudo-state filter error process~$\ef$,
\begin{small}
\begin{align*}
\ef &= 1_N \!\otimes\! \by_i - \yf \\
&= \left( 1_N \!\otimes\! \by_i - \ypp \right) - B^{\mathcal{C}}_i \left( 1_N \!\otimes\! \by_i - \ypp\right) \!-\! B^{\mathcal{I}}_i D_{H}^TR^{-1}\br_i \\
& \qquad - B^{\mathcal{I}}_i \widetilde{D}_H \left(1_N \!\otimes\! \by_i - \ypp\right) - B^{\mathcal{I}}_i\check{D}_H \left( 1_N \!\otimes\! \bx_i - \xpp \right) \\
&= \left( I_{\!M\!N} \!-\! B^{\mathcal{C}}_i \!-\! B^{\mathcal{I}}_i \!\widetilde{D}_{H}\right) \epp \!-\! B^{\mathcal{I}}_i \check{D}_{H} \epspp \!-\! B^{\mathcal{I}}_i\! D_{H}^T R^{-1} r_i.
\end{align*}
\end{small}
The state filtering update~\eqref{eqn:CIKF_xf}, in vector form, is
\begin{align*}
\xf = \xpp + K_i  \left( \yf \!-\! \left(I_N\!\otimes\!G\right) \xpp \right) ,
\end{align*}
where,~$K_i = \text{blockdiag} \{ K^1_i, \cdots, K^N_i \}$. Using the relation $\yf = \left(1_N\otimes\by_i\right) - \ef = \left(I_N\!\otimes\!G\right)\left(1_N\otimes\bx_i\right) - \ef $, we expand the state filter error process~$\ef$,
\begin{small}
\begin{align*}
\epsf &= 1_N \!\otimes\! \bx_i - \xf \\
&= \left( 1_{\!N} \!\otimes\! \bx_i - \xpp \right) \!-\! K_i \left(I_{\!N}\!\otimes\!G\right) \left(1_{\!N} \!\otimes\! \bx_i \!-\! \xpp\right) \!+\! K_i \ef \\
&= \left( I_{MN} - K^{\mathcal{C}}_i - K_i \left(I_N\!\otimes\!G\right) \right) \epspp + K_i \ef.
\end{align*}
\end{small}
The dynamics of the pseudo-state and state prediction errors,
\begin{align*}
\ep &= 1_N \!\otimes\! \by_{i+1} - \yp \\
&= \left(I_N\!\otimes\!\widetilde{A}\right)\left(1_N \!\otimes\!\by_i\right) + \left(I_N\!\otimes\! \check{A} \right)\left(1_N \!\otimes\!\bx_i\right) \\
&\quad + 1_N\!\otimes\!(G\bv_i) - \left(I_N\!\otimes\!\widetilde{A}\right)\yf -\left(I_N\!\otimes\! \check{A} \right)\xf \\
&= \left(I_N\!\otimes\!\widetilde{A}\right)\ef +  \left(I_N\!\otimes\! \check{A} \right) \epsf + 1_N\!\otimes\!(G\bv_i) \\
\epsp &= 1_N \!\otimes\! \bx_{i+1} - \xp \\
&= \left(I_N\!\otimes\!A\right)\left(1_N \!\otimes\!\bx_i\right) + 1_N\!\otimes\!\bv_i - \left(I_N\!\otimes\!A\right)\xf \\
&= \left(I_N\!\otimes\!A\right)\epsf + 1_N\!\otimes\!\bv_i.
\end{align*}
Since the state~$\bx_i$, pseudo-state~$\by_i$, their initial condition and all the noises are Gaussian, their estimates are also Gaussian making all the filtering and prediction errors Gaussian. \qed


\subsection{Proof of Lemma \ref{lemma:noises}}
\label{app:proof_noises}
Lemma~\ref{lemma:error_dynamics} and Assumption~\ref{assm:gauss} guarantee that $\epsf, \epsp, \ef, \bv_i, \br_i$ are Gaussian. The error noises~$\widetilde{\bm \phi}_i$ and~${\bm \phi}_i$ are therefore Gaussian as they are linear combinations of the error processes and the model noises $\epsf, \epsp, \ef, \bv_i, \br_i$. By Lemma~\ref{lemma:error_unbiased}, we have $\mathbb{E}\left[\epsf\right] = \mathbb{E}\left[\epsp\right] = \mathbb{E}\left[\ef\right] = \bm{0}$. From Assumption~\ref{assm:gauss}, we know $\mathbb{E}\left[\bv_i\right] = \mathbb{E}\left[\br_i\right] = \bm{0}$. We take expectation on both sides of~\eqref{eqn:til_phi}-\eqref{eqn:phi} and apply these relations

\begin{small}
\begin{align*}
\mathbb{E}\left[\widetilde{\bm \phi}_i\right] &= \!\left( I_{\!N} \!\otimes\! \check{A}\right)\! \mathbb{E}\left[\epsf\right] - \left(I_{\!N} \!\otimes\! \widetilde{A}\right)\! B^{\mathcal{I}}_i D_{H}^T R^{-1} \mathbb{E}\left[r_i\right]  \nonumber \\
& \quad + 1_{\!N} \!\otimes\! \left(G \mathbb{E}\left[\bv_i\right]\right) - \left(I_{\!N} \!\otimes\! \widetilde{A}\right)\!B^{\mathcal{I}}_i \check{D}_{H} \mathbb{E}\left[\epspp\right] = \bm{0}, \\
\mathbb{E}\left[{\bm \phi}_i\right] &= \left( I_{\!N} \!\otimes\! A\right)\! K_i \mathbb{E}\left[\ef\right] + 1_N \otimes \mathbb{E}\left[\bv_i\right] = \bm{0}.
\end{align*}
\end{small}
Combining \eqref{eqn:error_dynamics_epsf}, \eqref{eqn:error_dynamics_ep} and \eqref{eqn:til_phi}, we have
\begin{small}
	\begin{align*}
	\widetilde{\bm \phi}_i &= F_1 \epspp + F_2 \epp - F_3 D_{H}^T R^{-1} r_i + 1_{\!N} \!\otimes\! \left(G \bv_i\right)
	\end{align*}
\end{small}
where,
\begin{small}
	\begin{align}
	\label{eqn:F1}
	F_1 &= \left( I_{\!N} \!\otimes\! \check{A}\!\right)\! \left( I_{\!M\!N} \!-\! K_i \!\left(I_{\!N}\!\otimes\!G\right)\! -\! K_i \!B^{\mathcal{I}}_i \check{D}_{H}\!\! \right) \!-\! \left( I_{\!N} \!\otimes\! \widetilde{A}\!\right)\! \!B^{\mathcal{I}}_i \check{D}_{H} \\
	\label{eqn:F2}
	F_2 &= \left( I_{\!N} \!\otimes\! \check{A}\!\right)\! K_i \left( I_{\!M\!N} \!-\! B^{\mathcal{C}}_i \!-\! B^{\mathcal{I}}_i  \! \widetilde{D}_{H} \right) \\
	\label{eqn:F3}
	F_3 &= \left( I_{\!N} \!\otimes\! \widetilde{A}\!\right)\! \!B^{\mathcal{I}}_i.
	\end{align}
\end{small}
Since $\widetilde{\bm \phi}_i$ and~${\bm \phi}_i$ are zero-mean, the noise covariances are,
\begin{small}
	\begin{align}
	\widetilde{\Phi}_i &\!=\! \mathbb{E} \! \left[\!\widetilde{\bm \phi}_i \widetilde{\bm \phi}^T_i \! \right] \!=\! F_1 \Sigma_{i|i-1} F_1^T \!+\! F_2 P_{i|i-1} F_2^T \!+\! F_3 \overline{D}_H F^T_3 \nonumber \\
	\label{eqn:til_Phi_dynamics}
	& \qquad \qquad + J \!\otimes\! \left(GVG\right) + F_1 \Pi_{i|i-1} F_2^T + F_2 \Pi_{i|i-1}^T F_1^T,  \\
	\label{eqn:Phi_dynamics}
	\Phi_i &\!=\! \mathbb{E}\! \left[ \! {\bm \phi}_i {\bm \phi}^T_i \! \right] \!=\! \left( I_{\!N} \!\otimes\! A\right)\! K_i \Sigma_{i|i} \! K^T_i \! \left( I_{\!N} \!\otimes\! A^T\right) \!+\! J \!\otimes\! V,
	\end{align}
\end{small}
where, $F_1, F_2$ and $F_3$ are defined in~\eqref{eqn:F1}-\eqref{eqn:F3}. \qed
%


\subsection{Proof of Lemma \ref{lemma:uncorrelation}}
\label{app:proof_uncorrelation}
We first compute the conditional means~$\overline{\widetilde{\bm \theta}}^n_i$ and~$\overline{\bm \theta}^n_i$ of the new information~$\widetilde{\bm \theta}^n_i$ and~${\bm \theta}^n_i$ from~\eqref{eqn:new_info}. The means~$\overline{\widetilde{\bm \theta}}^n_i$ and~$\overline{\bm \theta}^n_i$ depend on the conditional means of~$\ypp^{l}, \widetilde{\bm z}^n_{i-1}$ and~$\yf^n$.
\begin{small}
\begin{align*}
&\mathbb{E} \begin{bmatrix} \ypp^{l} | \widetilde{\bm z}^n_{i-1}, \{ \widehat{\bm y}^l_{i-1|i-2} \}_{l \in \Omega_n} \end{bmatrix} \\
&= \mathbb{E} \left[ \mathbb{E} \left[ \by_i | \widetilde{\bm z}^l_{i-1}, \{ \widehat{\bm y}^k_{i-1|i-2} \}_{k \in \Omega_l} \right] \;|\; \widetilde{\bm z}^n_{i-1}, \{ \widehat{\bm y}^l_{i-1|i-2} \}_{l \in \Omega_n} \right] \\
&= \mathbb{E} \left[ \mathbb{E} \left[ \by_i | \widetilde{\bm z}^n_{i-1}, \{ \widehat{\bm y}^l_{i-1|i-2} \}_{l \in \Omega_n} \right] \;|\; \widetilde{\bm z}^l_{i-1}, \{ \widehat{\bm y}^k_{i-1|i-2} \}_{k \in \Omega_l} \right] \\
&= \mathbb{E} \left[ \ypp^{n} \;|\; \widetilde{\bm z}^l_{i-1}, \{ \widehat{\bm y}^k_{i-1|i-2} \}_{k \in \Omega_l} \right] = \ypp^{n}, \quad \forall\; l \in \Omega_n. \\
&\mathbb{E} \begin{bmatrix} \widetilde{\bm z}^n_i \;|\; \widetilde{\bm z}^n_{i-1}, \{ \widehat{\bm y}^l_{i-1|i-2} \}_{l \in \Omega_n} \end{bmatrix} \\
&= \mathbb{E} \begin{bmatrix} \widetilde{H}_n\by_i + H^T_n R_n^{-1} \br^n_i + \check{H}_n  \bx_i \;|\; \widetilde{\bm z}^n_{i-1}, \{ \widehat{\bm y}^l_{i-1|i-2} \}_{l \in \Omega_n} \end{bmatrix} \\
&= \widetilde{H}_n \mathbb{E} \begin{bmatrix} \by_i \;|\; \widetilde{\bm z}^n_{i-1}, \{ \widehat{\bm y}^l_{i-1|i-2} \}_{l \in \Omega_n} \end{bmatrix} \\
& \qquad \qquad \qquad + \check{H}_n \mathbb{E} \begin{bmatrix} \bx_i \;|\; \widetilde{\bm z}^n_{i-1}, \{ \widehat{\bm y}^l_{i-1|i-2} \}_{l \in \Omega_n} \end{bmatrix} \\
&= \widetilde{H}_n \ypp^n \!+\! \check{H}_n \mathbb{E} \left[ \mathbb{E} \left[ \bx_i | \yf^n, \widehat{\bm x}^l_{i-1|i-2}, \widetilde{\bm z}^n_{i-1}, \widehat{\bm y}^l_{i-1|i-2}, \in \Omega_n \right] \right]  \\
&= \widetilde{H}_n \ypp^n + \check{H}_n \mathbb{E} \left[ \xpp^n \;|\; \widetilde{\bm z}^n_{i-1}, \{ \widehat{\bm y}^l_{i-1|i-2} \}_{l \in \Omega_n}\right]  \\
&= \widetilde{H}_n \ypp^n + \check{H}_n \xpp^n. \\
&\mathbb{E} \begin{bmatrix} \yf^n | \widehat{\by}_{i-1|i-1}^n \end{bmatrix} = \mathbb{E} \left[ \by_i - \ef^n \;|\; \widehat{\by}_{i-1|i-1}^n \right] \\
&= \mathbb{E} \left[ \by_i \;|\; \widehat{\by}_{i-1|i-1}^n \right] - \mathbb{E} \left[\ef^n \;|\; \widehat{\by}_{i-1|i-1}^n \right] \\
&= G \mathbb{E} \left[ \bx_i \;|\; \widehat{\by}_{i-1|i-1}^n \right] = G\xpp^n.
\end{align*}
\end{small}
The conditional means of~$\ypp^{l}, \widetilde{\bm z}^n_{i-1}$ and~$\yf^n$ shows that the new uncorrelated information defined in~\eqref{eqn:nu1_tilde}-\eqref{eqn:nu1} expands to the vectors~$\widetilde{\bm \nu}^n_i$ and~${\bm \nu}^n_i$ in~\eqref{eqn:nu}. Note that~$\widetilde{\bm \nu}^n_i$ and~${\bm \nu}^n_i$ are zero mean, by definition, and, are Gaussian since they are linear combination of Gaussian vectors. Now to prove~$\widetilde{\bm \nu}^n_i$ and~${\bm \nu}^n_i$ are sequence of uncorrelated vectors, we have to show:
\begin{align*}
\mathbb{E}\left[ \widetilde{\bm \nu}^n_i \widetilde{\bm \nu}^{n^T}_j\right] = \mathbb{E}\left[ {\bm \nu}^n_i {\bm \nu}^{n^T}_j\right] = {\bm 0}, \; \forall \; i\neq j, \; \forall \; n.
\end{align*}
First, we write~$\widetilde{\bm \nu}^n_i$ and~${\bm \nu}^n_i$ in terms of the filtering and prediction error processes using~\eqref{eqn:pstate_obs}, \eqref{eqn:efn}-\eqref{eqn:epspn},
\begin{small}
\begin{align*}
	\widetilde{\bm \nu}^n_i \!=\! \begin{bmatrix}\epp^n \!-\! \epp^{l_1} \\ \vdots \\ \epp^n \!-\! \epp^{l_{d_n}} \\ \widetilde{H}_{\!n} \epp^n \!+\! \check{H}_{\!n} \epspp^n + H^T_n R^{-1}_n \br^n_i\end{bmatrix}\!\!\!, \;\;
	{\bm \nu}^n_i \!=\!  G\epspp^n \!-\! \ef^n.
\end{align*}
\end{small}
Without loss of generality, we consider~$i > j$. The rest of the proof is similar to the proof of Lemma~$5$ in~\cite{das2015TSP}. Here, the only difference is that we should condition on~$\{\widetilde{\bm z}^l_{i-1}\}_{l \in \Omega_n}, \{ \widehat{\bm y}^l_{i-1|i-1} \}_{l \in \Omega_n}$. \qed

\begin{center}
	{\bf Proof of Theorems}
	\vspace{-8pt}
\end{center}


\subsection{Proof of Theorem \ref{thm:OKDF}}
\label{app:proof_OKDF}
In Lemma~\ref{lemma:uncorrelation}, we showed that~$\widetilde{\bm \nu}^n_i$ and~${\bm \nu}^n_i$ are independent Gaussian sequences. By the Innovations Property~\cite{anderson2012optimal}, there are $1-1$ correspondence between~$\{ \widetilde{\bz}^n_i, \{ \ypp^l \}_{l \in \overline{\Omega}_n} \}$ and ~$\widetilde{\bm \nu}^n_i$, and between~$\{ \yf^n \}$ and~${\bm \nu}^n_i$. The Innovations Property guarantees that there exists a unique way to get one from the other.
\begin{small}
\begin{align*}
\yf^n \!=\! \mathbb{E} \!\left[ \by_i \;|\; \widetilde{\bz}^n_i, \{ \ypp^l \}_{l \in \overline{\Omega}_n} \right] ~&\Longleftrightarrow~ \yf^n \!=\! \mathbb{E} \!\left[ \by_i \;|\; \widetilde{\bm \nu}^n_i \right] \\
\xf^n \!=\! \mathbb{E} \left[ \bx_i \;|\; \yf^n \right] ~&\Longleftrightarrow~ \xf^n \!=\! \mathbb{E} \left[ \bx_i \;|\; {\bm \nu}^n_i \right]
\end{align*}
\end{small}
By the Gauss-Markov principle,
\begin{align*}	
\yf^n &= \ypp^n + \widehat{B}^n_i \widetilde{\bm \nu}^n_i \\
\xf^n &= \xpp^n + K^n_i {\bm \nu}^n_i
\end{align*}
where $\widehat{B}^n_i$ are the non-zero blocks of the~$n^{\text{th}}$ row of~$B_i$. Now we expand the term~$\widehat{B}^n_i \widetilde{\bm \nu}^n_i$ by multiplying the gain blocks~$B^{nl}_i$ with the corresponding~$(\ypp^l - \ypp^n)$ and the gain block~$B^{nn}_i$ with~$\left( \widetilde{\bz}^n_i \!\!-\! \widetilde{H}_{\!n} \ypp^n \!-\! \check{H}_{\!n} \xpp^n \right)$. This gives us the {\it consensus$+$innovations} filtering pseudo-state update~\eqref{eqn:CIKF_yf}.

The pseudo-state and state prediction updates are
\begin{align*}
\yp^n &\!=\! \mathbb{E} \left[ \by_{i+1} \;|\; \widetilde{\bz}^n_i, \{ \ypp^l \}_{l \in \overline{\Omega}_n} \right] \\
&\!=\! \mathbb{E} \left[ \widetilde{A}\by_i + G\bv_i + \check{A}\bx_i \;|\; \widetilde{\bz}^n_i, \{ \ypp^l \}_{l \in \overline{\Omega}_n} \right] \\
&\!=\! \widetilde{A} \yf^n + \check{A} \xf^n \\
\xp^n &\!=\! \mathbb{E} \left[ \bx_{i\!+\!1} \;|\; \yf^n  \right] \!=\! \mathbb{E} \left[ A \bx_i + \bv_i \;|\; \yf^n \right] = A \xf^n. \quad \qed
\end{align*}
%


\subsection{Proof of Theorem \ref{thm:covariance}}
\label{app:proof_covariance}
By Lemma~\ref{lemma:error_unbiased} and Lemma~\ref{lemma:error_dynamics}, the error processes, $\ef$, $\ep$, $\epsf$, and $\epsp$ are zero-mean Gaussian. The Lyapunov-type iterations~\eqref{eqn:Pf_dynamics}-\eqref{eqn:Pip_dynamics} of the filter and predictor error covariances, $P_{i|i}$, $\Sigma_{i|i}$, $\Pi_{i|i}$, $P_{i+1|i}$, $\Sigma_{i+1|i}$, and $\Pi_{i+1|i}$, follow directly from the definitions~\eqref{eqn:Pf}-\eqref{eqn:Pip} and error dynamics~\eqref{eqn:error_dynamics_ef}-\eqref{eqn:error_dynamics_epsp} by algebraic manipulations. \qed


\subsection{Proof of Theorem \ref{thm:capacity}}
\label{app:proof_capacity}
For any square matrix, $\rho(\widetilde{F}) \leq \| \widetilde{F} \|$. Hence if $\| \widetilde{F} \| < 1$, then it implies that $\rho(\widetilde{F}) < 1$. We derive the tracking capacity with the sufficient condition, $\| \widetilde{F} \| < 1$,
\begin{align*}
\| \widetilde{F} \| &= \begin{Vmatrix} \left( I_{\!N} \!\otimes\!\widetilde{A}\!\right)\! \left( I_{\!M\!N} \!-\! B^{\mathcal{C}}_i \!-\! B^{\mathcal{I}}_i  \! \widetilde{D}_{H} \right) \end{Vmatrix}  \\
&\leq  \begin{Vmatrix} I_{\!N} \!\otimes\!\widetilde{A}\! \end{Vmatrix} \begin{Vmatrix} \left( I_{\!M\!N} \!-\! B^{\mathcal{C}}_i \!-\! B^{\mathcal{I}}_i  \! \widetilde{D}_{H} \right) \end{Vmatrix} \\
&\leq  \begin{Vmatrix} GAG^{\dagger}\! \end{Vmatrix} \begin{Vmatrix} \left( I_{\!M\!N} \!-\! B^{\mathcal{C}}_i \!-\! B^{\mathcal{I}}_i  \! \widetilde{D}_{H} \right) \end{Vmatrix} \\
&\leq \frac{\lambda_m}{\lambda_1} \begin{Vmatrix} A\! \end{Vmatrix} \begin{Vmatrix} \left( I_{\!M\!N} \!-\! B^{\mathcal{C}}_i \!-\! B^{\mathcal{I}}_i  \! \widetilde{D}_{H} \right) \end{Vmatrix},
\end{align*}
where, $\| G \|_2 = \lambda_m$ and $\| G^{\dagger} \|_2 = \frac{1}{\lambda_1}$. Since $G$ is a symmetric positive semi-definite matrix, its spectral norm is its largest eigenvalue~$(\lambda_m)$ and the spectral norm of its pseudo-inverse,~$G^{\dagger}$, is the inverse of its smallest non-zero eigenvalue~$(\lambda_1)$. If there exists~$B^{\mathcal{C}}_i$ and~$B^{\mathcal{I}}_i$ such that
\begin{align*}
\frac{\lambda_m}{\lambda_1} \begin{Vmatrix} A\! \end{Vmatrix} \begin{Vmatrix} \left( I_{\!M\!N} \!-\! B^{\mathcal{C}}_i \!-\! B^{\mathcal{I}}_i  \! \widetilde{D}_{H} \right) \end{Vmatrix} < 1,
\end{align*}
then $\| \widetilde{F} \| < 1$ and also $\rho(\widetilde{F}) < 1$. The bound on the spectral norm of~$A$ is,
\begin{align*}
\| A \| &< \frac{\lambda_1}{\lambda_m  \begin{Vmatrix} \left( I_{\!M\!N} \!-\! B^{\mathcal{C}}_i \!-\! B^{\mathcal{I}}_i  \! \widetilde{D}_{H} \right) \end{Vmatrix} } \\
&\leq \max_{B^{\mathcal{C}}_i, B^{\mathcal{I}}_i} \frac{\lambda_1}{\lambda_m  \begin{Vmatrix} \left( I_{\!M\!N} \!-\! B^{\mathcal{C}}_i \!-\! B^{\mathcal{I}}_i  \! \widetilde{D}_{H} \right) \end{Vmatrix} } = C.
\end{align*}
Thus as long as $\| A \|_2 < C$, there exists~$B^{\mathcal{C}}_i$ and~$B^{\mathcal{I}}_i$ such that~$\rho(\widetilde{F}) < 1$. By global detectability Assumption~\ref{assm:detectability}, there exists~$K_i$ such that~$\rho(F) < 1$. Refer to~\cite{anderson2012optimal}, for the convergence conditions of the centralized information filters. Further, by Lemma~\ref{lemma:noises} the Gaussian noises processes~$\widetilde{\bm \phi}_i$ and~${\bm \phi}_i$ have bounded noise covariances. Thus, if $\| A \|_2 < C$, then from~\eqref{eqn:ep_stability}-\eqref{eqn:epsp_stability} we conclude that the CIKF~\eqref{eqn:CIKF_yf}-\eqref{eqn:CIKF_xp} converges with bounded MSE.  \qed


\subsection{Proof of Theorem \ref{thm:gain}}
\label{app:proof_gain}
By the Innovations Property and the Gauss-Markov principle~\cite{anderson2012optimal}, the optimal gains~$\widehat{B}^n_i$ and~$K^n_i$ in~\eqref{eqn:CIKF_yf_nu}-\eqref{eqn:CIKF_xf_nu} are:
\begin{align*}
	\widehat{B}^n_i &= \Sigma_{\by_i \widetilde{\bm \nu}^n_i} \left(\Sigma_{\widetilde{\bm \nu}^n_i}\right)^{-1}, \\
	K^n_i &= \Sigma_{\bx_i {\bm \nu}^n_i} \left(\Sigma_{{\bm \nu}^n_i}\right)^{-1}
\end{align*}
that yield minimized MSE estimates~$\yf^n$ and~$\xf^n$ of the pseudo-state~$\by_i$ and of the field~$\bx_i$, respectively, at each agent~$n$. The cross-covariances~$\Sigma_{\by_i \widetilde{\bm \nu}^n_i}$ and~$\Sigma_{\bx_i {\bm \nu}^n_i}$ are
\begin{small}
\begin{align*}
&\Sigma_{\by_i \widetilde{\bm \nu}^n_i} = \mathbb{E} \left[ \left(\by_i - \overline{\by}_i \right) \widetilde{\bm \nu}^{n^T}_i \right] \\
&= \mathbb{E} \!\left[ \!\left(\epp^n \!+\!\left(\ypp^n \!-\! \overline{\by}_i \right) \right)\!\! \begin{bmatrix} \epp^n \!-\! \epp^{l_1} \\ \vdots \\ \epp^n \!-\! \epp^{l_{d_n}} \\ \widetilde{H}_{\!n} \epp^n \!\!+\! \check{H}_{\!n} \epspp^n \!+\! H^T_n R^{-1}_n \br^n_i \end{bmatrix}^{\!\!T} \!\right] \\
&= \mathbb{E}\! \left[ \epp^n \! \left[ \!\!
\begin{array}{l;{2pt/2pt}l;{1pt/1pt}l;{2pt/2pt}l}
\!\epp^{n^T} \!-\! \epp^{l^T_1}\! &\! \cdots\! &\! \epp^{n^T} \!-\! \epp^{l^T_{d_n}} \!&\! \widetilde{H}_{\!n} \epp^{n^T} \\
& &  & \; + \check{H}_{\!n}\epspp^{n^T} \!\!
\end{array}
\!\! \right] \! \right] \\
&= \left[
\begin{array}{l;{2pt/2pt}l;{1pt/1pt}l;{2pt/2pt}l}
\!\!\!\! P_{i|i\!-\!1}^{nn} \!-\! P_{i|i\!-\!1}^{nl_1} & \cdots & P_{i|i\!-\!1}^{nn} \!-\! P_{i|i\!-\!1}^{nl_{d_n}} & P_{i|i\!-\!1}^{nn} \! \widetilde{H}^T_n \!+\! \Pi_{i|i\!-\!1}^{nn^T} \! \check{H}^T_n \!\!\!\!
\end{array}
\right], \\
&\Sigma_{\bx_i {\bm \nu}^n_i} = \mathbb{E} \left[ \left(\bx_i - \overline{\bx}_i \right) {\bm \nu}^{n^T}_i \right] \\
&= \mathbb{E} \left[ \left(\epspp^n +\left(\xpp^n - \overline{\bx}_i \right) \right) \left( G\epspp^{n} \!-\! \ef^{n}  \right)^T \right] \\
&= \mathbb{E} \left[ \epspp^n \! \left( G\epspp^{n} \!-\! \ef^{n}  \right)^T \! \right] = \Sigma_{i|i\!-\!1}^{nn} G - \Gamma^{nn}_i.
\end{align*}
\end{small}

In the above derivation, using the iterated law of expectation it can be shown that the terms~$\mathbb{E} \left[ \left(\ypp^n - \overline{\by}_i \right) \widetilde{\bm \nu}^{n^T}_i \right] = \mathbb{E} \left[ \left(\xpp^n - \overline{\bx}_i \right) {\bm \nu}^{n^T}_i \right] = {\bm 0}$. Also, $\mathbb{E} \left[ \epp^n \br^{n^T}_i \right] = {\bm 0}$ due to the statistical independence of the noise sequences. Similarly, the covariances matrices~$\Sigma_{\widetilde{\bm \nu}^n_i}$ and~$\Sigma_{{\bm \nu}^n_i}$ are:
\begin{small}
	\begin{align*}
	&\Sigma_{\widetilde{\bm \nu}^n_i} = \mathbb{E} \left[ \widetilde{\bm \nu}^{n}_i \widetilde{\bm \nu}^{n^T}_i \right] \\
	&= \mathbb{E} \left[ \begin{bmatrix} \epp^n \!-\! \epp^{l_1} \\ \vdots \\ \epp^n \!-\! \epp^{l_{d_n}} \\ \widetilde{H}_{\!n} \epp^n \!+\! \check{H}_{\!n} \epspp^n \\ \qquad + H^T_n R^{-1}_n \br^n_i \end{bmatrix} \begin{bmatrix} \epp^n \!-\! \epp^{l_1} \\ \vdots \\ \epp^n \!-\! \epp^{l_{d_n}} \\ \widetilde{H}_{\!n} \epp^n \!+\! \check{H}_{\!n} \epspp^n \\ \qquad + H^T_n R^{-1}_n \br^n_i \end{bmatrix}^T \right] \\
	&\Sigma_{{\bm \nu}^n_i} = \mathbb{E} \left[ {\bm \nu}^{n}_i {\bm \nu}^{n^T}_i \right] = \mathbb{E} \left[ \left( G\epspp^n \!-\! \ef^n \right) \left( G\epspp^n \!-\! \ef^n \right)^T \right].
	\end{align*}
\end{small}
The rest of the derivation of~$\Sigma_{\widetilde{\bm \nu}^n_i}$ and~$\Sigma_{{\bm \nu}^n_i}$ in terms of the error covariance matrices is by block-by-block multiplication of the above expressions. \qed


\bibliographystyle{IEEEtran}
\bibliography{DoubleColumn_DasMoura_TSP2016}

\begin{thebibliography}{10}
\providecommand{\url}[1]{#1}
\csname url@samestyle\endcsname
\providecommand{\newblock}{\relax}
\providecommand{\bibinfo}[2]{#2}
\providecommand{\BIBentrySTDinterwordspacing}{\spaceskip=0pt\relax}
\providecommand{\BIBentryALTinterwordstretchfactor}{4}
\providecommand{\BIBentryALTinterwordspacing}{\spaceskip=\fontdimen2\font plus
\BIBentryALTinterwordstretchfactor\fontdimen3\font minus
  \fontdimen4\font\relax}
\providecommand{\BIBforeignlanguage}[2]{{%
\expandafter\ifx\csname l@#1\endcsname\relax
\typeout{** WARNING: IEEEtran.bst: No hyphenation pattern has been}%
\typeout{** loaded for the language `#1'. Using the pattern for}%
\typeout{** the default language instead.}%
\else
\language=\csname l@#1\endcsname
\fi
#2}}
\providecommand{\BIBdecl}{\relax}
\BIBdecl

\bibitem{kalman1961new}
R.~E. Kalman and R.~S. Bucy, ``New results in linear filtering and prediction
  theory,'' \emph{Journal of basic Engineering}, vol.~83, no.~3, pp. 95--108,
  1961.

\bibitem{kalman1960new}
R.~E. Kalman, ``A new approach to linear filtering and prediction problems,''
  \emph{Journal of Basic Engineering}, vol.~82, no.~1, pp. 35--45, 1960.

\bibitem{de2011kalman}
J.~R.~P. de~Carvalho, E.~D. Assad, and H.~S. Pinto, ``Kalman filter and
  correction of the temperatures estimated by precis model,'' \emph{Atmospheric
  Research}, vol. 102, no.~1, pp. 218--226, 2011.

\bibitem{battistelli2015distributed}
G.~Battistelli, L.~Chisci, N.~Forti, G.~Pelosi, and S.~Selleri, ``Distributed
  finite element {K}alman filter,'' in \emph{European Control
  Conference}.\hskip 1em plus 0.5em minus 0.4em\relax IEEE, 2015, pp.
  3695--3700.

\bibitem{das2015TSP}
S.~Das and J.~M.~F. Moura, ``Distributed {K}alman filtering with dynamic
  observations consensus,'' \emph{IEEE Transactions on Signal Processing},
  vol.~63, no.~17, pp. 4458--4473, 2015.

\bibitem{khan2010connectivity}
U.~A. Khan, S.~Kar, A.~Jadbabaie, and J.~M.~F. Moura, ``On connectivity,
  observability, and stability in distributed estimation,'' in \emph{49th IEEE
  Conference on Decision and Control}, 2010, pp. 6639--6644.

\bibitem{olfati2005distributed}
R.~Olfati-Saber, ``Distributed {K}alman filter with embedded consensus
  filters,'' in \emph{44th IEEE Conference on Decision and Control, and 8th
  European Control Conference}, 2005, pp. 8179--8184.

\bibitem{olfati2007distributed}
R.~Olfati-Saber, ``Distributed {K}alman filtering for sensor networks,'' in
  \emph{46th IEEE Conference on Decision and Control}, 2007, pp. 5492--5498.

\bibitem{khan2008distributing}
U.~A. Khan and J.~M.~F. Moura, ``Distributing the {K}alman filter for
  large-scale systems,'' \emph{IEEE Transactions on Signal Processing},
  vol.~56, no.~10, pp. 4919--4935, 2008.

\bibitem{carli2008distributed}
R.~Carli, A.~Chiuso, L.~Schenato, and S.~Zampieri, ``Distributed {K}alman
  filtering based on consensus strategies,'' \emph{IEEE Journal on Selected
  Areas in Communications}, vol.~26, no.~4, pp. 622--633, 2008.

\bibitem{schizas2008consensus}
I.~D. Schizas, G.~B. Giannakis, S.~I. Roumeliotis, and A.~Ribeiro, ``Consensus
  in ad hoc {WSN}s with noisy links - {Part II}: Distributed estimation and
  smoothing of random signals,'' \emph{IEEE Transactions on Signal Processing},
  vol.~56, no.~4, pp. 1650--1666, 2008.

\bibitem{ribeiro2010kalman}
A.~Ribeiro, I.~D. Schizas, S.~Roumeliotis, and G.~B. Giannakis, ``{K}alman
  filtering in wireless sensor networks,'' \emph{IEEE Control Systems
  Magazine}, vol.~30, no.~2, pp. 66--86, 2010.

\bibitem{olfati2009kalman}
R.~Olfati-Saber, ``{K}alman-consensus filter: Optimality, stability, and
  performance,'' in \emph{48th IEEE Conference on Decision and Control, and
  28th Chinese Control Conference}, 2009, pp. 7036--7042.

\bibitem{cattivelli2010diffusion}
F.~S. Cattivelli and A.~H. Sayed, ``Diffusion strategies for distributed
  {K}alman filtering and smoothing,'' \emph{IEEE Transactions on Automatic
  Control}, vol.~55, no.~9, pp. 2069--2084, 2010.

\bibitem{casbeer2009distributed}
D.~W. Casbeer and R.~Beard, ``Distributed information filtering using consensus
  filters,'' in \emph{American Control Conference}.\hskip 1em plus 0.5em minus
  0.4em\relax IEEE, 2009, pp. 1882--1887.

\bibitem{khan2011networked}
U.~A. Khan and A.~Jadbabaie, ``Networked estimation under information
  constraints,'' \emph{arXiv preprint arXiv:1111.4580}, 2011.

\bibitem{khan2011coordinated}
U.~A. Khan and A.~Jadbabaie, ``Coordinated networked estimation strategies
  using structured systems theory,'' in \emph{50th IEEE Conference on Decision
  and Control, and 11th European Control Conference}, 2011, pp. 2112--2117.

\bibitem{khan2013genericity}
M.~Doostmohammadian and U.~A. Khan, ``On the genericity properties in
  distributed estimation: Topology design and sensor placement,'' \emph{IEEE
  Journal of Selected Topics in Signal Processing}, vol.~7, no.~2, pp.
  195--204, 2013.

\bibitem{kar2011gossip}
S.~Kar and J.~M.~F. Moura, ``Gossip and distributed {K}alman filtering: Weak
  consensus under weak detectability,'' \emph{IEEE Transactions on Signal
  Processing}, vol.~59, no.~4, pp. 1766--1784, 2011.

\bibitem{li2015distributed}
D.~Li, S.~Kar, J.~M.~F. Moura, H.~V. Poor, and S.~Cui, ``Distributed {K}alman
  filtering over massive data sets: analysis through large deviations of random
  {R}iccati equations,'' \emph{IEEE Transactions on Information Theory},
  vol.~61, no.~3, pp. 1351--1372, 2015.

\bibitem{kar2012distributed}
S.~Kar, J.~M.~F. Moura, and K.~Ramanan, ``Distributed parameter estimation in
  sensor networks: Nonlinear observation models and imperfect communication,''
  \emph{IEEE Transactions on Information Theory}, vol.~58, no.~6, pp.
  3575--3605, 2012.

\bibitem{kar2011convergence}
S.~Kar and J.~M.~F. Moura, ``Convergence rate analysis of distributed gossip
  (linear parameter) estimation: Fundamental limits and tradeoffs,'' \emph{IEEE
  Journal of Selected Topics in Signal Processing}, vol.~5, no.~4, pp.
  674--690, 2011.

\bibitem{martins2012augmented}
S.~Park and N.~C. Martins, ``An augmented observer for the distributed
  estimation problem for {LTI} systems,'' in \emph{American Control
  Conference}, 2012, pp. 6775--6780.

\bibitem{karmoura-spm2013}
S.~Kar and J.~M.~F. Moura, ``Consensus+innovations distributed inference over
  networks: cooperation and sensing in networked systems,'' \emph{IEEE Signal
  Processing Magazine}, vol.~30, no.~3, pp. 99--109, May 2013.

\bibitem{rao1991fully}
B.~S. Rao and H.~F. Durrant-Whyte, ``Fully decentralised algorithm for
  multisensor {K}alman filtering,'' in \emph{IEE Proceedings D - Control Theory
  and Applications}, vol. 138, no.~5.\hskip 1em plus 0.5em minus 0.4em\relax
  IET, 1991, pp. 413--420.

\bibitem{ribeiro2006soi}
A.~Ribeiro, G.~B. Giannakis, and S.~I. Roumeliotis, ``{SOI-KF}: Distributed
  {K}alman filtering with low-cost communications using the sign of
  innovations,'' \emph{IEEE Transactions on Signal Processing}, vol.~54,
  no.~12, pp. 4782--4795, 2006.

\bibitem{dimakiskarmourarabbatscaglione-2010}
A.~G. Dimakis, S.~Kar, J.~M.~F. Moura, M.~G. Rabbat, and A.~Scaglione, ``Gossip
  algorithms for distributed signal processing,'' \emph{Proceedings of the
  IEEE}, vol.~98, no.~11, pp. 1847--1864, Nov 2010.

\bibitem{das2013ICASSP}
S.~Das and J.~M.~F. Moura, ``Distributed state estimation in multi-agent
  networks,'' in \emph{38th IEEE International Conference on Acoustics, Speech
  and Signal Processing}, 2013, pp. 4246--4250.

\bibitem{das2013EUSIPCO}
S.~Das and J.~M.~F. Moura, ``Distributed {K}alman filtering,'' in \emph{21st
  European Signal Processing Conference}, 2013, pp. 1--5.

\bibitem{das2013Allerton}
S.~Das and J.~M.~F. Moura, ``Distributed linear estimationof dynamic random
  fileds,'' in \emph{51st Annual Allerton Conference on Communication, Control,
  and Computing}, 2013, pp. 1120--1125.

\bibitem{das2013Asilomar}
S.~Das and J.~M.~F. Moura, ``Distributed {K}alman filtering and network
  tracking capacity,'' in \emph{47th Asilomar Conference on Signals, Systems,
  and Computers}, 2013, pp. 629--633.

\bibitem{kar2014asymptotically}
S.~Kar and J.~M.~F. Moura, ``Asymptotically efficient distributed estimation
  with exponential family statistics,'' \emph{IEEE Transactions on Information
  Theory}, vol.~60, no.~8, pp. 4811--4831, 2014.

\bibitem{mohammadi2015distributed}
A.~Mohammadi and A.~Asif, ``Distributed consensus innovation particle filtering
  for bearing/range tracking with communication constraints,'' \emph{IEEE
  Transactions on Signal Processing}, vol.~63, no.~3, pp. 620--635, 2015.

\bibitem{chung1997spectral}
F.~R. Chung, \emph{Spectral Graph Teory}.\hskip 1em plus 0.5em minus
  0.4em\relax American Mathematical Society, 1997, vol.~92.

\bibitem{anderson2012optimal}
B.~D. Anderson and J.~B. Moore, \emph{Optimal filtering}.\hskip 1em plus 0.5em
  minus 0.4em\relax Courier Dover Publications, 2012.

\bibitem{shahrampour2016distributed}
S.~Shahrampour, A.~Rakhlin, and A.~Jadbabaie, ``Distributed estimation of
  dynamic parameters: Regret analysis,'' \emph{arXiv preprint
  arXiv:1603.00576}, 2016.

\end{thebibliography}

\vskip-20pt
\begin{IEEEbiography}[{\includegraphics[width=1in,height=1.25in,clip,keepaspectratio]{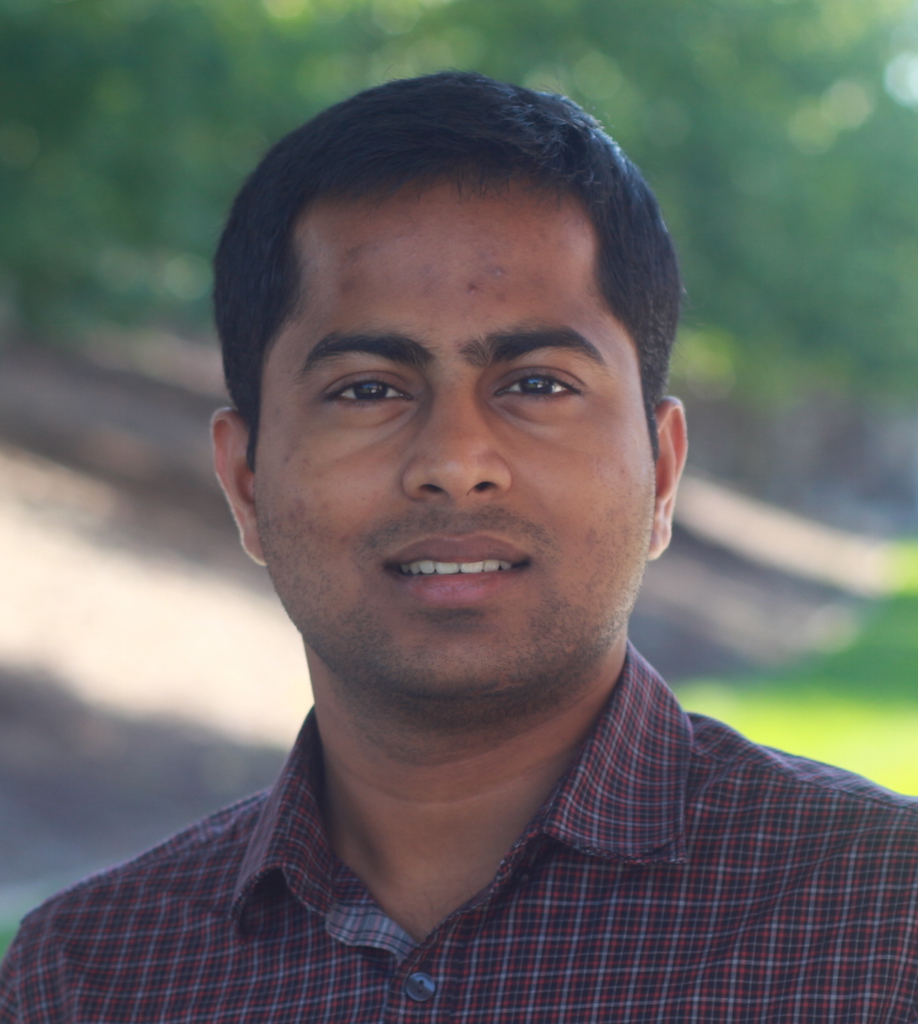}}]{Subhro~Das}(S'12--M'16) received the Bachelor of Technology (B.Tech.) degree in Electronics and Electrical Communication Engineering from Indian Institute of Technology, Kharagpur, India, in~2011, and the MS and PhD degrees in Electrical and Computer Engineering from Carnegie Mellon University, Pittsburgh PA, in 2014 and 2016, respectively. 
	
Since July 2016, he is a Postdoctoral Researcher at IBM T.J. Watson Research Center, Yorktown Heights, NY. His research interests are in statistical inference, distributed estimation over multi-agent networks, model adaptation in dynamic environments and time-series analysis; broadly in the areas of statistical signal processing, machine learning and data-mining with applications in healthcare and smart infrastructure.
\end{IEEEbiography}

\vskip-25pt

\begin{IEEEbiography}[{\includegraphics[width=1in,height=1.25in,clip,keepaspectratio]{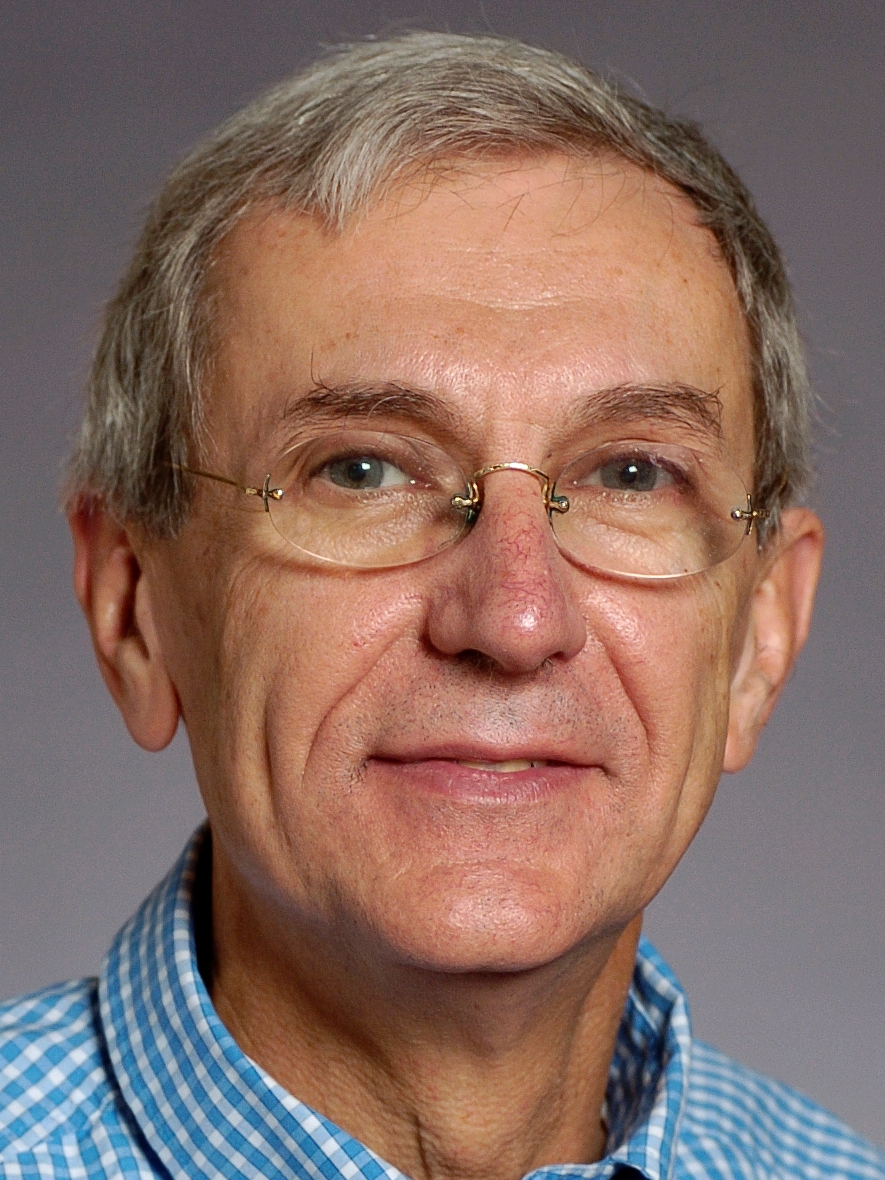}}]{Jos\'e M.~F.~Moura}(S'71--M'75--SM'90--F'94) received the engenheiro electrot\'{e}cnico degree from Instituto Superior T\'ecnico (IST), Lisbon, Portugal, and the M.Sc., E.E., and D.Sc.~degrees in EECS from the Massachusetts Institue of Technology (MIT), Cambridge, MA.
	
	He is the Philip L.~and Marsha Dowd University Professor at Carnegie Mellon University (CMU). He was on the faculty at IST and has held visiting faculty appointments at MIT and New York University (NYU). He founded and directs a large education and research program between CMU and Portugal, www.icti.cmu.edu.
	
	His research interests are on  data science, graph signal processing, and statistical and algebraic signal and image processing. He has published over 550 papers and holds thirteen patents issued by the US Patent Office. The technology of two of his patents (co-inventor A. Kav\v{c}i\'c) are in about three billion disk drives read channel chips of 60~\% of all computers sold in the last 13 years worldwide and was, in 2016, the subject of the largest university verdict/settlement in the information technologies area.
	
	Dr. Moura is the IEEE Technical Activities Vice-President (2016) and member of the IEEE Board of Directors. He served in several other capacities including IEEE Division IX Director, member of several IEEE Boards, President of the IEEE Signal Processing Society(SPS), Editor in Chief for the IEEE Transactions in Signal Processing, interim Editor in Chief for the IEEE Signal Processing Letters.
	
	Dr. Moura has received several awards, including  the Technical Achievement Award and the Society Award from the IEEE Signal Processing. In 2016, he received the CMU College of Engineering Distinguished Professor of Engineering Award. He is a Fellow of the IEEE, a Fellow of the American Association for the Advancement of Science (AAAS), a corresponding member of the Academy of Sciences of Portugal, Fellow of the US National Academy of Inventors, and a member of the US National Academy of Engineering.
\end{IEEEbiography}

\end{document}